\documentclass[10pt,journal,compsoc]{IEEEtran}
\usepackage{amsmath,amsfonts}
\usepackage{algorithmic}
\usepackage{array}
\usepackage{booktabs} % nicer tables
\usepackage{tabularx} % also nicer tables
\usepackage[caption=false,font=normalsize,labelfont=sf,textfont=sf]{subfig}
\usepackage{textcomp}
\usepackage{stfloats}
\usepackage{cite}
\usepackage[hyphens]{url}
\usepackage{verbatim}
\usepackage{graphicx}
\def\BibTeX{{\rm B\kern-.05em{\sc i\kern-.025em b}\kern-.08em
    T\kern-.1667em\lower.7ex\hbox{E}\kern-.125emX}}
\usepackage{balance}
\usepackage{hyperref}

\usepackage{fp} % arithmetics for colorboxes
\usepackage{ifthen} % conditions for colorboxes
\usepackage{xcolor}

\usepackage{cleveref}
\usepackage{enumitem}

\def\NAT@spacechar{~}% NEW

\newcommand{\ie}{i.e., }
\newcommand{\eg}{e.g., }
\newcommand{\etal}[1]{{#1}~et~al.}

\newcommand{\textQuote}[1]{``\emph{#1}''}
\newcommand*\elide{\textup{[\,\dots]}}

\newcommand{\hui}{HUI} 
\newcommand{\huisubset}{MR-HUI}

\newcommand{\component}{IO component}

\newcommand{\components}{IO components}
\newcommand{\Components}{IO components}

\crefname{section}{Sec.}{Sec.}
\crefname{figure}{Fig.}{Fig.}
\crefname{table}{Tab.}{Tab.}

\newcommand{\lastaccess}{2025-12-01}

\newcommand{\descSpacing}{\hspace{0.25em}$\triangleright$}

\newcommand\ro[1]{{\smallskip\noindent\textbf{#1}\quad}}

\usepackage[normalem]{ulem}
\usepackage{soul}

\providecommand{\remove}[1]{\ignorespaces}
\providecommand{\add}[1]{{#1}}
\providecommand{\replace}[2]{#2}
\providecommand{\softremove}[1]{\ignorespaces}

\definecolor{myboxcolor}{HTML}{81a1c1}

\newcommand{\numberbox}[1]{%
    \FPdiv{\result}{#1}{1.16}%
    \FPadd{\result}{\result}{10}%
    \FPmin{\result}{\result}{116}%
    \colorbox{myboxcolor!\result}{\footnotesize\textcolor{black}{#1}}%
}

\newcommand{\numberboxexample}[1]{%
    \colorbox{myboxcolor}{\footnotesize\textcolor{black}{#1}}%
}

\newcommand{\surveyWebsite}[0]{\href{https://imldresden.github.io/huis}{\mbox{https://imldresden.github.io/huis}}}

\usepackage{orcidlink}
\author{%
    Sebastian Hubenschmid*~\orcidlink{0000-0002-8704-8503},
    Marc Satkowski*~\orcidlink{0000-0002-1952-8302},
    Johannes Zagermann*~\orcidlink{0000-0002-6757-9753},
    Juli\'{a}n M\'{e}ndez*~\orcidlink{0000-0003-1029-7656},\\
    Niklas Elmqvist~\orcidlink{0000-0001-5805-5301},
    Steven Feiner~\orcidlink{0000-0001-9978-7090},
    Tiare Feuchtner~\orcidlink{0000-0002-9922-5538},
    Jens Emil Gr{\o}nb{\ae}k~\orcidlink{0000-0002-9566-7195},
    Benjamin Lee~\orcidlink{0000-0002-1171-4741},\\
    Dieter~Schmalstieg~\orcidlink{0000-0003-2813-2235},
    Raimund~Dachselt~\orcidlink{0000-0002-2176-876X},
    and Harald Reiterer~\orcidlink{0000-0001-8528-8928}
    \thanks{*Authors contributed equally.}%
    \thanks{S. Hubenschmid, J. Zagermann, T. Feuchtner, and H. Reiterer are with the HCI Group, University of Konstanz, Germany.}%
    \thanks{M. Satkowski is with Fraunhofer Institute for Process Engineering and Packaging IVV, Dresden, Germany}
    \thanks{
    J. M\'{e}ndez and R. Dachselt are with the Interactive Media Lab Dresden at TUD Dresden University of Technology, Germany. R. Dachselt is also with the Centre for Scalable Data Analytics and Artificial Intelligence (ScaDS.AI) and the Centre for Tactile Internet (CeTI) at TUD Dresden University of Technology, Germany.}%
    \thanks{N. Elmqvist and J.E. Gr{\o}nb{\ae}k are with Aarhus University, Denmark.}%
    \thanks{Steven Feiner is with Columbia University, USA.}%
    \thanks{Benjamin Lee is with University of Stuttgart, Germany.}%
    \thanks{Dieter Schmalstieg is with Graz University of Technology, Austria and University of Stuttgart, Germany.}%
}

\title{Hybrid User Interfaces: Past,~Present,~and~Future~of~Complementary Cross-Device Interaction in Mixed Reality}

\IEEEtitleabstractindextext{%
\begin{abstract}

We investigate hybrid user interfaces (HUIs), aiming to establish a cohesive understanding and to adopt consistent terminology for this nascent research area.
HUIs combine heterogeneous devices in complementary roles, leveraging the distinct benefits of each.
Our work focuses on cross-device interaction between 2D devices and mixed reality environments, which are particularly compelling, leveraging the familiarity of traditional 2D platforms while providing spatial awareness and immersion.
Although prior work has prominently explored such \hui{}s in the context of mixed reality, we still lack a cohesive understanding of the unique design possibilities and challenges of such combinations, resulting in a fragmented research landscape.
We conducted a systematic survey and present a taxonomy of HUIs that combine conventional display technology and mixed reality environments.
Based on this, we discuss past and current challenges, the evolution of definitions, and prospective opportunities to tie together the past 30 years of research with our vision of future HUIs.

\end{abstract}

}

\begin{document}

\maketitle
\IEEEpubid{%
\begin{minipage}{\textwidth}
\centering
This work has been submitted to the IEEE for possible publication. Copyright may be transferred without notice, after which this version may no longer be accessible.
\end{minipage}
}

% !TEX root = ../main.tex

\section{Introduction}

% ===================
% v2.0
% ===================
\IEEEPARstart{I}{mmersive} augmented reality~(AR) and virtual reality~(VR) is gradually gaining relevance in everyday life, with affordable off-the-shelf hardware becoming increasingly available to consumers.
These mixed reality~(MR) platforms\footnote{In this work, mixed reality refers to both AR and VR (cf.~\cite{speicher2019what}), as discussed in \cref{sec:taxonomy:edge_cases}.} have now evolved to the point where researchers can explore the nuances of interaction design without being constrained by major technological limitations.
In this context, Feiner and Shamash~\cite{feiner1991hybrid} proposed in 1991 the concept of \textit{hybrid user interfaces} (\hui{}s), which combine \textQuote{heterogeneous display and interaction device technologies.}
This concept theoretically allows for an infinite integration of technologies, yet, such heterogeneous combinations are especially compelling for MR environments:
The ubiquity, convenience, and familiarity of conventional 2D platforms (\eg smartphones, desktops) provide a perfect complement to the immersion and complexity of optical see-through~(OST) and video see-through~(VST) head-worn devices~(HWDs)%
\footnote{We refer to \textQuote{head-worn devices}~(HWDs) instead of \textQuote{head-mounted displays}~(HMDs) to emphasize the increase in wearability and capabilities of current hardware, but intentionally kept the previous term HMD for describing older hardware.}.
Yet, although the concept of \hui{}s has persisted for decades, no coherent delineation has emerged.
As a result, there is a distinct lack of consistent design models and terminologies---fragmenting the research community across overlapping research areas such as cross-reality systems~\cite{auda2023scoping}, transitional interfaces~\cite{grasset2006transitional}, or cross-device interaction~\cite{brudy2019crossdevice}.

For example, the seminal cross-device taxonomy by Brudy~et~al.~\cite{brudy2019crossdevice} provides an overarching model for the research area of \hui{}s. 
However, their taxonomy only considers established devices (\eg mobile devices) and homogeneous combinations, leaving topics such as MR environments largely for future work.
Yet, without firm design principles and in the face of countless possible device combinations, the design space of \hui{}s can appear bewilderingly large.
Therefore, we extend the existing taxonomy by examining complementary cross-device interaction between traditional 2D device technologies and novel MR platforms as a prevalent subset of \hui{}s.
By first examining this subset, we take a first step towards a better understanding of the broader research area of \hui{}s.
Thus, we aim to establish a common terminology, allowing researchers and practitioners to better benefit from shared insights, establish a consistent framework, and inspire future systems.
This would, in turn, enable the creation of a cohesive understanding of the unique design possibilities and challenges of \hui{}s.

Toward this goal, we examine the past, present, and future of HUIs, focusing on the unique combination of MR environments with conventional display technologies, thus integrating the research area of \hui{}s into the wider cross-device taxonomy.
Overall, our work contributes:
\begin{itemize}
    \item A \textbf{positioning of \add{MR-}\hui{}s}, creating a better understanding of the term within the current fragmented research landscape (\cref{sec:rw} and \cref{sec:outlook}).
    
    \item A \textbf{systematic literature survey} (\cref{sec:methodology}) of \hui{}s that combine 2D devices and MR environments, from which we \textbf{present our own taxonomy} (\cref{sec:taxonomy}) and \textbf{identify current trends} (\cref{sec:results}).
    
    \item A \textbf{discussion of the challenges and research opportunities} that describe how our investigated \hui{}s have evolved (\cref{sec:discussion}), paving the way for their \textbf{future development and research} beyond current technological restrictions (\cref{sec:outlook}).
\end{itemize}

\section{What is a Hybrid User Interface?}
\label{sec:rw}

% >>> Intro

Since 1991, the term \textQuote{hybrid user interface} has continued to evolve.
Despite significant advances in device technologies, no coherent definition has emerged that delineates \hui{}s from related research areas.
%such as cross-device interaction.
This lack of common understanding makes it difficult to compare results and share insights.
% >>> Content
In the following, we first review prior definitions and the \textit{background} (\cref{sec:rw:background}) of \hui{}s, building on previous work by Satkowski~and~Méndez~\cite{satkowski2023fantastic}, and position \hui{}s in relation to current \textit{adjacent research areas} (\cref{sec:rw:adjacent}).
Based on this, we discuss how we can \textit{identify} \hui{}s (\cref{sec:rw:definition}) and derive three \textit{attributes} that specifically characterize  HUIs that combine MR environments with conventional display technologies (\cref{sec:rw:aspects}) to guide our literature survey.
%We then position \hui{}s in relation to current \textit{adjacent research areas} (\cref{sec:rw:adjacent}), provide a \textit{working definition} of \hui{}s and its subset (\cref{sec:rw:definition}), based on which we derive three \textit{attributes} (\cref{sec:rw:aspects}) to guide our survey.

% We then position \hui{}s in relation to current \textit{adjacent research areas} (\cref{sec:rw:adjacent}) and discuss how we are \textit{defining a hybrid user interface} (\cref{sec:rw:aspects}) in the context of this work.

% establish a common \textit{definition} (\ref{sec:rw:aspects}) for \hui{}s within the context of this work.

% Traditional computing devices (\eg desktops and mobile devices) are often combined to form one coherent system.
% Consequently, there is growing interest in combining such devices with mixed reality.
%introduce our interpretation of the term ``hybrid user interfaces'' (\cref{sec:rw:aspects}).
%extracting \textit{key aspects} (\cref{sec:rw:aspects}) of \hui{}s.

% #####################################################################
% ######## Other Content ##############################################
\begin{figure*}
  \centering
  \includegraphics[width=\textwidth]{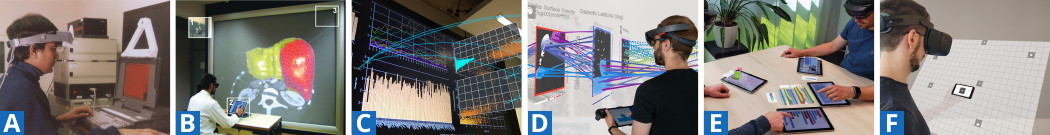}
  \caption{
    We take a look at 30 years of \textit{hybrid user interface} research and its evolution, focusing on combinations of conventional 2D devices with mixed reality environments. From left to right, 
    (A)~starting with its definition~\cite{feiner1991hybrid}, 
    (B)~early extensions~\cite{bornik2006hybrid} (Image \copyright~IEEE 2006), 
    and (C)~current usage for augmenting displays~\cite{reipschlager2021personal} (Image \copyright~\etal{Reipschläger}, CC BY 4.0), 
    (D)~extending tablets in single-user~\cite{hubenschmid2021stream} (Image \copyright~\etal{Hubenschmid}, CC BY 4.0)
    and (E)~multi-user scenarios~\cite{langner2021marvis} (Image \copyright~\etal{Langner}, CC BY 4.0),
    and (F)~finally shifting towards more empirical studies~\cite{hubenschmid2023smartphone} (Image \copyright~\etal{Hubenschmid}, CC BY 4.0). All images used with permission.
    To explore our investigated corpus, please visit \surveyWebsite.
  }
  % \Description{
  %     An image strip showing six different hybrid user interfaces.
  %     The first picture shows a person wearing a head-mounted display in front of a laptop.
  %     The second picture shows a person sitting at a table in front of a large projector displaying a 3D model.
  %     The third picture shows a large wall display with virtual elements and lines extending from the display in augmented reality.
  %     The fourth picture shows a person holding a tablet and wearing an augmented reality head-worn device, standing in front of virtual scatter plots.
  %     The fifth picture shows two persons wearing augmented reality head-worn devices and sitting at a desk with multiple tablets in front of them.
  %     Between each tablet, virtual content is displayed in augmented reality.
  %     The sixth picture shows a person wearing an augmented reality head-worn device and holding a smartphone.
  %     The smartphone is virtually extended with a semi-transparent screen-aligned display.
  % }
  \label{fig:teaser}
\end{figure*}

% #####################################################################
% !TEX root = ../main.tex

% ##########################################################
% ######## Sub-Section #####################################
% ##########################################################
\subsection{Background}
\label{sec:rw:background}

% >>> Intro
The term \hui{} was coined in the early 1990s by Feiner~and~Shamash, describing a combination of \textQuote{heterogeneous display and interaction device technologies}~\cite{feiner1991hybrid}.
% >>> Original definition
They argue that physical device sizes decreased with the advent of portable computing devices (\ie laptops). 
This led to reduced interface real estate while still retaining high-resolution input and output.
In contrast, immersive technologies such as MR (especially HWDs)---back then and still mostly today---offer a lower resolution for both input and output space. 
However, AR has the potential for virtually unlimited interfaces that exceed the capabilities of conventional display technologies.
%Thus, Feiner~and~Shamash propose to combine these two technologies by \textQuote{taking advantage of the strong points of each}~\cite{feiner1991hybrid}, treating the technologies as complementary instead of competing.
Thus, they propose to combine these technologies by \textQuote{taking advantage of the strong points of each}~\cite{feiner1991hybrid}, treating the technologies as complementary instead of competing.
They exemplify this concept in a ``hybrid window manager'', combining a high-resolution yet size-restricted desktop with a low-resolution yet virtually unlimited AR head-mounted display interface into a unified application that blurs the boundaries between interfaces.

This initially broad scope opened up a vast design space, but its ambiguity may have limited the adoption of the term within the field, leading to the diffusion of this term as it was applied to an increasing variety of device combinations over the years.
%utility in guiding research within the field.
%giving researchers ample room for interpretation,
% >>> The term over the years 
Although one prominent theme was the combination of 2D devices and immersive MR environments, this was not universally shared by all prior works.
% >>> 1
On the one hand, Butz~et~al. highlighted that \hui{}s extend to various \textQuote{technologies and techniques, including virtual elements such as 3D widgets, and physical objects such as tracked displays and input devices}~\cite{butz1999enveloping}.
% >>> 2
They noted that the resulting global AR space can be shared, which is also discussed by Feiner, as \hui{}s combine all devices \textQuote{in a mobile, shared environment}~\cite{feiner2000environment}.
% >>> 3
%The definition of \citet{bornik2006hybrid} adds another goal of \hui{}s: to \textQuote{pair 3D perception and direct 3D interaction with 2D system control and precise 2D interaction}.
Bornik~et~al. further emphasize the potential combination of MR environments with conventional 2D display technology for \hui{}s, as they \textQuote{pair 3D perception and direct 3D interaction with 2D system control and precise 2D interaction}~\cite{bornik2006hybrid}.
% >>> 5
This is echoed by Geiger~et~al., who state that \hui{}s \textQuote{combine 2D, 3D, and real object interaction and may use multiple input and output devices and different modalities}~\cite{geiger2008hyui}.
% >>> 6
In contrast, Strawhacker~and~Bers employ \hui{}s in a broader context without MR, presenting a \hui{} that allows \textQuote{users [to] switch freely between tangible and graphical input}~\cite{strawhacker2015want}.
%Concerning real objects, \citet{strawhacker2015want} present a \hui{} in which they combine a graphical and tangible user interface. 
% They highlight that \textQuote{users [should be able to] switch freely between tangible and graphical input}, whereof the former relates to wooden blocks.

% >>> 3
One commonly shared theme is the importance of \textit{complementarity}:
For example, Sandor~et~al. state that \textQuote{information [in \hui{}s] can be spread over a variety of different, but complementary, displays}~\cite{sandor2005immersive}.
Furthermore, in line with Butz~et~al.~\cite{butz1999enveloping}, they describe that users of \hui{}s can \textQuote{interact through a wide range of interaction devices}~\cite{sandor2005immersive}---we thus see the potential of \hui{}s not within a random assortment of technologies, but in a principled integration of different standalone \textQuote{interaction devices}.

\begin{table*}[h!]
\centering
\caption{\add{An overview of adjacent research areas and their relation to \hui{}s.}}
\renewcommand{\arraystretch}{1.2} % Row height
% \rowcolors{2}{gray!10}{white} % Alternating row colors

\begin{tabularx}{\linewidth}{
    >{\raggedright\arraybackslash}l
    % >{\arraybackslash}l
    >{\raggedright\arraybackslash}p{5cm}
    >{\raggedright\arraybackslash}X
}

% \rowcolor{headerblue!80} % Header background color
\textbf{Research Area} &
% \textbf{Amount of HUIs} &
% \textbf{Reference} &
\textbf{Description} &
\textbf{Relation to HUIs}\\

\midrule
\textbf{Distributed User Interfaces}~\cite{elmqvist2011distributed}  &
% 5000 &
Foundation for all research on multi-device usage. &
\hui{}s are one possible realization, focusing on distributing interfaces across heterogeneous devices.\\

\textbf{Complementary Interfaces}~\cite{zagermann2022complementary} &
% 100000 &
Symbiotic combinations of devices and interfaces through complementarity. & 
\hui{}s focus on complementarity between devices and thus are a technological approach to complementary interfaces.\\

\textbf{Cross-Device Interaction}~\cite{brudy2019crossdevice} &
% 100000 &
Describes general research on homogeneous and heterogeneous device combinations. &
\hui{}s are a subset of cross-device interaction focusing on heterogeneous combinations.\\

\textbf{Cross-Reality Environments}~\cite{auda2023scoping} &
% 100000 &
Systems and interfaces that span across different points on the reality-virtuality continuum. &
Significant amounts of prior work in \hui{}s combine MR environments (\eg AR and VR) with conventional displays (\ie reality), thus representing a subset of cross-reality.\\

\textbf{Transitional Interfaces} \cite{mayer2024crossing} &
% No reference paper &
Describes transitions between environments, interfaces, and devices. &
HUIs may use transitional interfaces when switching between devices; likewise, transitional interfaces may encompass multiple heterogeneous devices to be switched between.\\

\bottomrule
\end{tabularx}
\label{table:adjacent}
\end{table*}

        %Our survey focuses on the specific subset of \textit{hybrid user interfaces} that uses a heterogeneous \textit{cross-reality environment} (\ie{} combining 2D \todo{devices?} with MR environments).
        % Such todo{\textit{hybrid augmented interfaces} are thus located at the intersection between \textit{hybrid user interfaces} and \textit{cross-reality environments}.
        % Yet, there are also potential \textit{cross-device}, \textit{cross-reality} systems that fall outside of our definition of \textit{hybrid user interfaces}, such as combining multiple AR and VR HWDs.
        % \todo{maybe leave this out}: While these terms describe technical realizations, the areas of \textit{complementary interfaces} and \textit{distributed user interfaces} describe overarching conceptual areas that span across a subset of these terms.

% !TEX root = ../main.tex

% ##########################################################
% ######## Sub-Section #####################################
% ##########################################################
\subsection{Adjacent Research Areas}
\label{sec:rw:adjacent}

The combination of complementary device technologies opens up a vast design space with countless possible combinations.
These device combinations are, however, not exclusive to \hui{}s. 
%As a result, different terminology is used to describe similar concepts, while similar terminologies may refer to different concepts.
In this section, we describe the intersection between \hui{}s and adjacent research and clearly position \hui{}s within the current research landscape \add{(see also \cref{table:adjacent} for an overview)}.
\subsubsection{Distributed User Interfaces}

Elmqvist~\cite{elmqvist2011distributed} defines the term \textit{distributed user interface} as an interface whose components are distributed across one or more dimensions, such as input, output, platform, space, and time.
This theoretical perspective can be seen as the foundation of any research on multi-device usage, including multimodal interaction.
For \hui{}s, all proposed dimensions are relevant, as they span multiple technologies and distribute the user interface accordingly.
However, by combining heterogeneous technologies, \hui{}s aim to distribute input and output to the most suitable device to achieve a goal while focusing on their complementary use.
Therefore, we see \textit{distributed user interfaces} as an overarching conceptual model, with \hui{}s representing a possible technical realization of this concept that focuses on heterogeneous devices.

% ##########################################################
% ######## Paragraph #######################################
% ##########################################################
\subsubsection{Complementary Interfaces}

%\citet{feiner1991hybrid} suggested to treat multiple technologies in a complementary way, thereby taking advantage of their individual benefits.
%This can be seen as one of the key aspects of \hui{}s, by \eg complementing 2D visual and interaction spaces with MR environments.
Recent works by Zagermann~et~al.~\cite{zagermann2022complementary} as well as Elmqvist~\cite{elmqvist2023anywhere} highlight that attributing unique roles, properties, and purposes to individual devices and modalities can lead to meaningful combinations of interfaces that support users in their current task at hand.
%Following \citeauthor{elmqvist2011distributed}'s~\cite{elmqvist2011distributed} dimensions of distributed user interfaces, 
Such \textit{complementary interfaces} distribute interaction across devices and modalities to establish a \textQuote{symbiosis of interfaces, where each component purposefully increases the quality of interaction}~\cite{zagermann2022complementary}.
In the context of \hui{}s, complementary interfaces can be considered as an overarching concept that includes combinations of homogeneous and heterogeneous devices, but also input (\eg interaction techniques) and output modalities (\eg visually or auditory).
The core ideas of complementary interfaces are typically part of \hui{}s, in that \hui{}s represent a technical realization of this concept.
%We see \hui{}s at the intersection between cross-device interaction and complementary interfaces.

% Complementary Interfaces can be seen as an overarching concept of how to design multi-device and multimodal experiences.
% Its core ideas are typically part of \hui{}s; however, research on \hui{}s mostly focuses on the combination of multiple components (\ie devices) to take advantage of the strong points of each.
% Yet, the concept of Complementary Interface could help to pave the way for the future of \hui{}s:
% By redefining what constitutes a component beyond a type of device, future research can create meaningful combinations of a variety of devices, but also with respect to specific input and output modalities and capabilities.
% We will further discuss the future of \hui{}s in Section~\ref{sec:outlook}.

%Recent works by Zagermann and Hubenschmid et al.~\cite{zagermann2022complementary} as well as Elmqvist~\cite{elmqvist2023anywhere} highlight...
%- One aspect commonly mentioned throughout hybrid UIs is complementary
%- see Figure~\ref{fig:venn}

% ##########################################################
% ######## Paragraph #######################################
% ##########################################################
\subsubsection{Cross-Device Interaction}

Brudy et al.~\cite{brudy2019crossdevice} provide a comprehensive overview of the field of \textit{cross-device interaction}.
Here, the focus is on research that \textQuote{transcends the individual device and user}~\cite{brudy2019crossdevice}, which unifies research that is focused on different kinds of multi-device environments, ranging from multi-monitor setups to ad-hoc mobile device ecologies.
%Cross-device interaction is often related to Weiser's vision of ubiquitous computing~\cite{weiser1999computer}, studying devices such as smartphones, tablets, or larger interactive surfaces.
Although Brudy et al.~\cite{brudy2019crossdevice} list head-worn MR devices and tangibles as part of their cross-device taxonomy, they do not further elaborate on combinations of heterogeneous (\eg non-immersive and immersive) devices. 

Therefore, cross-device interaction can be seen as an umbrella term that includes research not only on homogeneous but also heterogeneous combinations---the latter of which includes \hui{}s.
Some of the research on homogeneous cross-device interaction (\eg attention switching) can be transferred to \hui{}s.
However, \hui{}s have unique challenges and opportunities, for example, with regard to heterogeneous roles of devices, conflicting interaction spaces, and co-dependencies of devices.

%We thereby not only extend \citeauthor{brudy2019crossdevice}'s \cite{brudy2019crossdevice} taxonomy, but also unify multi-device research in a broader sense.

%- Most closely aligned to hybrid UIs is the research area of cross-device interaction, which \textQuote{definition...}~\cite{brudy2019crossdevice}.
%- Survey by \citeauthor{brudy2019crossdevice}~\cite{brudy2019crossdevice} shows [...]; includes distributed UIs~\cite{elmqvist2011distributed} [...]
%- In contrast to established cross-device interaction, hybrid UIs always contain mixed reality output component that are complemented with a 2D space, thereby creating unique challenges and opportunities
%- Still, many aspects of CDI are relevant to hybrid UIs, as 
%- We therefore see CDI as immediate umbrella term and extend/build upon existing survey by focusing on the AR part that was not considered
%- While we build on cross-device taxonomy, / instead of repeating content from cross-device interaction, we focus on particular case of hybrid UIs
  %  - we argue that a more narrow focus on this particular case of cross-device interaction can help ``provide guidance about scope and specializations of research within'' this field.

%- (maybe) problems with multi-device attention switching are still present / more pronounced due to additional switch in focal distance...

% ##########################################################
% ######## Paragraph #######################################
% ##########################################################
\subsubsection{Cross-Reality Environments}

Using the reality--virtuality (RV) continuum~\cite{milgram1994taxonomy} as a foundation, the research area of \textit{cross-reality environments} investigates the benefits of combining various points on the RV continuum.
Key aspects of cross-reality environments include \textQuote{smooth transition[s] between systems using different degrees of virtuality} and \textQuote{collaboration between users using different systems with different degrees of virtuality}~\cite{simeone2020international}.
Recent taxonomies by Auda~et~al.~\cite{auda2023scoping} and Wang~and~Maurer~\cite{wang2022design} further solidify concepts such as transitioning between different points on the RV continuum or concurrently using multiple distinct systems along the RV continuum.

For a single user, such cross-reality environments typically create a sequence of actions (\eg switching devices).
Depending on the workflow, this could be described as a migratory interface~\cite{brudy2019crossdevice}, asynchronous \hui{}~\cite{hubenschmid2021asynchronous}, or transitional interface~\cite{mayer2024crossing, carvalho2012design}.
Furthermore, given that previous work often focuses on \hui{}s that combine 2D \components{} with MR environments (\eg \cite{feiner1991hybrid, bornik2006hybrid, butz1999enveloping, geiger2008hyui}), we see cross-reality as a potential umbrella term, where a subset of prior \hui{}s (see \cref{sec:rw:definition}) lies at the intersection between cross-reality and cross-device interaction:
While cross-device interaction focuses on the combination of \textit{devices}, the area of cross-reality concentrates on the combination of \textit{realities}.
Thus, the area of cross-reality research encompasses both homogeneous (\eg collaboration between HWDs in different realities) and heterogeneous device combinations (\eg switching from desktop to VR). 
We consider the latter as part of our \hui{} terminology.

\subsubsection{Transitional Interfaces}

The MagicBook by Billinghurst~et~al.~\cite{billinghurst2001magicbook} is often described as the first \textit{transitional interface}~\cite{mayer2024crossing, carvalho2012design}: Here, a user can move seamlessly along the RV continuum---from browsing the physical book to a handheld AR display to immersive VR.
Although they can be regarded as a subset of cross-reality, transitional interfaces specifically focus on the design of transitions and their effect on users~\cite{mayer2024crossing}.
%Research has studied transitions (1)~from one manifestation of the RV continuum to another (\eg from AR to VR), (2)~along the entire RV continuum, and (3)~within individual manifestations (\eg from VR to VR).
%The latter can be seen as a clear distinction to cross-reality environments.
While early head-mounted displays required a discrete switch of hardware to move across interfaces, currently available HWDs allow for continuous transitions by interactively adding or removing virtual contents.

Similarly to research on cross-reality environments, transitional interfaces typically involve a sequence of activities:
A user solves one aspect of a given task in AR and continues to work on the activity in VR.
The transition keeps the user oriented in the task space, choosing one manifestation of the RV continuum at a time.
However, for \hui{}s, connecting to multiple manifestations at the same time is a key ability.
In line with previous research, we \textQuote{think of HUIs and transitional interfaces as complementary}~\cite{carvalho2012design}, rather than competing:
A \hui{} may use a transitional interface when switching between devices; likewise, a transitional interface may encompass multiple heterogeneous devices to be switched between.

%In the following, we will describe key aspects that allowed us to classify \hui{}s.
%Recent (mini?) survey: \cite{mayer2024crossing}
%carvalho2012design: \textit{``Altogether, they comprise a diversity of hardware and software settings that can be combined to create hybrid environments such as HUIs (Fig. 1). Actually, it makes sense to think of HUIs and transitional interfaces as complementary.''}

% ##########################################################
% ######## Paragraph #######################################
% ##########################################################

\subsubsection{Further Related Terms}

Adjacent research areas additionally use related terms to describe their work:
The term cross-device interaction is often used interchangeably with the term ``cross-surface interaction'', while earlier work in that area is often considered as ``multi-device'' or ``multi-display'' systems~\cite{brudy2019crossdevice}.
Similarly, there are other terms that are related to \hui{}s, but focus on interaction techniques (\ie ``hybrid interaction''~\cite{knierim2021smartphone, besancon2021state}) or the setting (\ie ``hybrid virtual environment''~\cite{carvalho2012design, decarvalho2009designing, wang2014coordinated}).
The term ``augmented displays'' was previously used to describe integral concepts of \hui{}s (cf. \cite{reipschlager2020augmented, reipschlager2018debugar}).
Therefore, we consider it as a unique configuration for \hui{}s.
%(see also \cref{sec:taxonomy}).

% \smallskip
% The described terms for adjacent research areas and related terms influenced the search strategy for relevant literature on \hui{}s as described in \cref{sec:prisma:strategy}.

\subsection[Characterizing Hybrid User Interfaces]{Characterizing Hybrid User Interfaces}
\label{sec:rw:definition}

Research on \hui{}s shares several common themes (\eg \textit{complementarity}, \textit{heterogeneity}), yet the term remains technology-driven and potentially misunderstood:
%, with no clear definition that delineates the broad scope of this term, which led to a fragmentation and misunderstanding of this term:
A large number of prior \hui{}s describe the term as a combination of 2D and 3D technology~\cite{bornik2006hybrid, elmqvist2011distributed, hinckley1994survey}, including the initial system demonstration by Feiner~and~Shamash~\cite{feiner1991hybrid}.
%as also presented in the initial definition of Feiner~and~Shamash~\cite{feiner1991hybrid}.
Moreover, the term \hui{} was previously also used for different constellations of \components{} (\eg desktop combined with tangibles~\cite{strawhacker2015want}, mobile devices~\cite{bessghaier2017usability}, or conversational interfaces~\cite{liu2022conversationbased}), indicating wider applicability.
% (cf. cross-device interaction~\cite{brudy2019crossdevice}).
For the purpose of our survey, we modernize the definition of \hui{}s by Feiner~and~Shamash~\cite{feiner1991hybrid} within the current research landscape:

\begin{quote}
    %\textit{Hybrid User Interfaces are an area of cross-device computing that leverages distinct benefits of heterogeneous \components{}.}
    \textit{Hybrid User Interfaces are an area of cross-device computing that leverages distinct benefits of heterogeneous components with input and output (\components{}).}
\end{quote}

%Yet, this seems incongruous with other \hui{}s that employ the term in different constellations 
%TODO: therefore, establishing a concrete definition may be hard and we instead acknowledge the vagueness of the term?
%TODO: further, many system that, based on these insights, we would identify as \hui{} are spread across adjacent research areas (thus not called \hui{}). -> This makes it hard to establish taxonomy of shared characteristics and identify relevant systems.

In contrast to its initial definition, we explicitly concentrate on conceptual device capabilities instead of specific technologies to avoid being limited by current hardware capabilities, but are intentionally vague about possible device combinations (\eg see cross-device interaction~\cite{brudy2019crossdevice} for a potential ontology).
%; we also discuss \hui{}s as a fuzzy concept in \cref{sec:discussion:definition}).
%We acknowledge that \hui{}s are a technology-driven research area with an ever-shifting window of opportunities due to available hardware capabilities (see \cref{sec:outlook}).
To this end, we use the term \textit{\component{}} throughout this work to refer to a standalone device set with the input and output capabilities necessary for interacting with a given application (\eg desktop with mouse and keyboard, VR HWD with controllers).

While this definition highlights the broad potential design space of \hui{}s, it also lacks specificity and, therefore, limits its utility for our survey:
For example, a combination of tangibles and a desktop computer might be considered a \hui{} and likewise, a combination of an AR HWD with a tablet---increasing the blurriness of the research stream instead of increasing its focus.
Therefore, to explore this vast design space and gain concrete insights into commonalities of \hui{}s, \textit{we focus our work on the specific \hui{} subset combining 2D with MR-enabled \components{}}.
This combination captures our interest, as it
    was initially presented by Feiner~and~Shamash~\cite{feiner1991hybrid}, 
    is most prevalent in prior \hui{} literature~\cite{satkowski2023fantastic}, 
    and also reflects on work presented at the IEEE ISMAR workshop on \hui{}s (2023)~\cite{hubenschmid2023hybrid}, constituting the most recent and comprehensive outlet for work on \hui{}s.
We thereby explicitly exclude the vast research area of \textit{tangible interaction}~\cite{ishi2003tangible} or nascent research areas (\eg brain--computer interfaces~\cite{chun2016bci}, Internet of Things~\cite{sanctorum2019unifying}) from this subset, thereby increasing the specificity of our survey.

For the purpose of differentiating between the results of our survey and possible implications to the broader field of \hui{}s as well as to improve the readability of our paper, we will refer to this subset as \textit{\huisubset{}}.
%, clarifying our focus on the prevalent association of \hui{}s with MR environments. 
%Given the already fragmented research landscape and infinite possible combinations of heterogeneous \components{}, we intend to use this term only within the scope of this paper.

%For the purpose of our survey, we call our subset \textit{\huisubset{}} to refer to the combination of conventional 2D display technology and MR environments.
%Given the already fragmented research landscape, we only intend to use this term to delineate our concrete results from the wider area of \hui{}s and encourage the use of the broader term ``\hui{}'' for future systems.

% !TEX root = ../main.tex

% ##########################################################
% ######## Sub-Section #####################################
% ##########################################################
%\subsection{Attributes of Our Subset of (?) Hybrid User Interfaces}
\subsection[Attributes of Mixed Reality Hybrid User Interfaces]{Attributes of Mixed Reality Hybrid User Interfaces}
\label{sec:rw:aspects}

We framed our work on the contemporary and most prevalent subset of \hui{}s: a combination of 2D \components{} with MR environments.
% We framed our work on \hui{}s that combine 2D \components{} with MR environments---reflecting a contemporary and prevalent subset.
%This view of \hui{}s does not consider key aspects, such as the relationship and dependency of these components.
We derive three \textbf{attributes} based on previous usage of the term to further guide our literature review.
Please note that the first two attributes are indicative of the broader area of \hui{}s, while the last attribute specifies the constraints of \huisubset{}s.
%To guide the process of our literature survey, we need 
%Based on prior usage of the term and their position within the current research landscape, we, therefore, derive three \textit{key qualities} of \hai{}s to guide our literature survey, focusing on their potential in combining MR environments with conventional 2D display technologies.

% ##########################################################
% ######## Paragraph #######################################
% ##########################################################
% \paragraph*{\textbf{Multiple \components{} in heterogeneous roles}}
\ro{\textbf{Multiple \components{} in heterogeneous roles.}}
\hui{}s typically combine multiple \components{} where each fulfills a need that is not adequately addressed by other components.
Although each \component{} is self-contained, a \hui{} is deliberately spread across multiple components to intentionally \textQuote{take advantage of the strong points of each}~\cite{feiner1991hybrid}.
%Unlike prior definitions that focus on combinations of heterogeneous \textit{technologies},
We also see the potential beyond technological aspects and broaden our scope to include combinations of heterogeneous \textit{roles}.

%We also consider homogeneous device combinations (e.g., handheld AR instead of AR HWDs) and instead focus on the heterogeneity of responsibilities within their device ecology: for example, we consider handheld AR through a smartphone combined with another smartphone as a potential \hui{} (cf.~e.g.,~\citep{vanukuru2020dual}).

% ##########################################################
% ######## Paragraph #######################################
% ##########################################################
% \paragraph*{\textbf{\Components{} are codependent, forming a single application}}
\ro{\textbf{\Components{} are codependent.}}
%5
A \hui{} is composed of multiple \components{}, but the real power does not emerge from any individual \component{}, but from the interaction of all of them~\cite{weiser1999computer}. 
A \hui{} thus acts as one holistic application from the user's perspective.
The deliberate spread of responsibilities across \components{} requires a certain degree of co-dependency between \components{} (\ie a system may become nonfunctional without all \components{} present).
%By offloading essential features to one component (\ie taking advantage of the strong points), the whole system may become inoperative without it.
We examine both synchronous (\ie using \components{} in parallel) and asynchronous usage (\ie using \components{} in sequence)~\cite{bornik2006hybrid, hubenschmid2021asynchronous}.

% ##########################################################
% ######## Paragraph #######################################
% ##########################################################
% \paragraph{\textbf{Complementing 2D visual and interaction spaces with mixed reality environments}}

% An essential aspect of \hui{}s is to \textQuote{pair 3D perception and direct 3D interaction with 2D system control and precise 2D interaction}~\cite{bornik2006hybrid}.
% Given the (at times) conflicting needs of 2D and 3D spaces, combining components that excel at either 2D or 3D interaction creates a system that compromises on neither but excels at both.
% It is not necessary to use the full 3D space:
% The unlimited visual and interaction space of MR environments can also be used to extend the capabilities of current technologies beyond their physical limit (e.g., seamless 2D display extension).

% ##########################################################
% ######## Paragraph #######################################
% ##########################################################
% \paragraph*{\textbf{Complementing 2D visual and interaction capabilities with mixed reality environments}}
\ro{\textbf{Complementing 2D with MR \components{}.}}
A common aspect found in prior \hui{} literature is to \textQuote{pair 3D perception and direct 3D interaction with 2D system control and precise 2D interaction}~\cite{bornik2006hybrid}.
%Here, we understand ``interaction space'' (cf.~\cite{feiner1991hybrid}) as [...].
Given the (at times) conflicting needs of 2D and 3D spaces, combining 2D and MR \components{} can yield a superior result.
%These advantages are not necessarily limited to 3D perception, but can also include other advantages, such as the virtually unlimited screen space of MR environments (cf. \cite{feiner1991hybrid}).
For the remainder of this work, we refer to such \components{} as either \textit{2D \component{}} or \textit{MR \component{}}.

\section{Review Methodology}
\label{sec:methodology}

% >>> Intro
We look at research that matches our previously specified attributes to obtain corpus of \huisubset{} publications.
This section describes the process for identifying, filtering, and analyzing relevant publications, following the PRISMA~\cite{page2021prisma} guidelines.
We present the general search strategy (\cref{sec:prisma:strategy}), selection process (\cref{sec:prisma:selection}), and data extraction (\cref{sec:prisma:extraction}). 
Moreover, the limitations of our procedure (\cref{sec:prisma:limitations}) and possible extensions of the survey (\cref{sec:prisma:extension}) are highlighted.
Finally, we describe how this survey can be used by other researchers and practitioners (\cref{sec:prisma:how-to-use}).
The complete survey corpus, its coding, and the scripts used to prepare and analyze the data can be found in the supplementary material.

% #####################################################################
% ######## Other Content ##############################################
% #####################################################################
% !TEX root = ../main.tex

% #      ##############################################################
% #      # Figure #####################################################
% #      ##############################################################
\begin{figure}[t!]
    \centering
    \includegraphics[width=.48\textwidth]{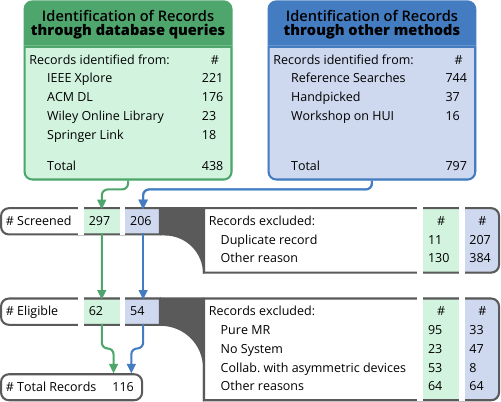}
    \caption{
        Overview of our reviewing process including paper counts, following PRISMA. 
        Records were rejected once a single exclusion criterium was fulfilled, yet, some records potentially fulfilled multiple criteria (\eg survey papers~\cite{auda2023scoping}).
        The initial corpus of papers (n=\add{1235}) including all codes is part of the supplemental material of this submission.
    }
    % \Description{
    %     Flowchart of our survey following PRISMA guidelines.
    %     It starts with two initial sets (one for database queries and one for other methods).
    %     The papers from these two sets are first screened and then checked for eligibility.
    %     The total number of records is 116.
    % }
    \label{fig:prisma}
\end{figure}

% !TEX root = ../main.tex

% ##########################################################
% ######## Sub-Section #####################################
% ##########################################################
\subsection{Search Strategy}
\label{sec:prisma:strategy}

% >>> Intro
In accordance with the PRISMA guidelines, we collected the papers using two different identification approaches (see \cref{fig:prisma}), which we present in this section.

% ##########################################################
% ######## SubSub-Section ##################################
% ##########################################################
\ro{Identification via databases.}
%\label{sec:prisma:strategy:db}
% >>> Intro
Our survey focuses on publications that combine 2D \components{} with MR environments.
As the term ``\hui{}'' is not consistently used in previous work, we used adjacent terms and their synonyms (see \cref{sec:rw:adjacent}) to build a query that captures a broader range of potential publications that present \hui{} applications.

% >>> Terms
We iteratively refined our keywords and evaluated them on a base corpus.
Starting from the initial keyword ``hybrid user interface'', we identified synonyms used in the literature and extended our set of keywords.
This resulted in three sets of keywords focused on different aspects that we aimed to capture.
First, \textit{Set A} includes keywords that could be used interchangeably with the term \hui{}.
Next, \textit{Set~B} consists of adjacent terms and their synonyms that are not necessarily MR-specific.
This is why, with \textit{Set C}, we further narrow down the search to MR-related terms.
The complete list of terms in each set is as follows\footnote{All terms were used in singular and plural, as well as with or without hyphens (where appropriate).}:

\begin{description}
    \item[Set A:] ``Hybrid User Interface'', ``Complementary Interface'', ``Augmented Display'', or ``Cross-Reality''
    \item[Set B:] ``Hybrid Virtual Environment'', ``Hybrid Interaction'', ``Cross-Device'', ``Cross-Surface'', ``Multi-Device'', ``Multi-Display'', ``Distributed User Interface'', or ``Transitional Interface'' 
    \item[Set C:] ``Augmented Reality'', ``Virtual Reality'', ``Mixed Reality'', or ``Extended Reality''
\end{description}

% >>> Query
The query was constructed using the expression: \verb +Set A or (Set B and Set C)+. 
    %OR( OR ( Set A ), AND( OR( Set B ), OR( Set C ) ) )
% >>> Library Selection
We focused on archival, peer-reviewed publications, including full and short papers in journals and conferences, as well as book chapters, workshop submissions, posters, and works in progress.
We searched in common digital libraries for publications in Human--Computer Interaction and Visualization, namely \href{https://ieeexplore.ieee.org/}{IEEE Xplore}, \href{https://dl.acm.org/}{ACM Digital Library}, \href{https://link.springer.com/}{Springer Link}, and \href{https://onlinelibrary.wiley.com/}{Wiley Online Library}.%\footnote{Links to the libraries: \href{https://ieeexplore.ieee.org/}{IEEE Xplore}, \href{https://dl.acm.org/}{ACM DL}, \href{https://onlinelibrary.wiley.com/}{Wiley Online Library}, \href{https://link.springer.com/}{Springer Link}}.
% >>> Results
The searches with the described query%
\footnote{The specific search syntax for each library can be found in the supplemental material.} 
in these libraries%
\footnote{%
    \add{The advanced search of the Springer library only allows searching in the title and keywords and not in the abstract.}
    } %
resulted in \add{438} publications (see \cref{fig:prisma}).
The cut-off date for all searches was \add{December 11th, 2025}.

% ##########################################################
% ######## SubSub-Section ##################################
% ##########################################################
\ro{Identification via other methods.}
%\label{sec:prisma:strategy:om}
% >>> Handpicked and Other sources
In addition to the query search\add{---and in line with PRISMA guidelines---}we selected papers through three other methods (see \cref{fig:prisma}).
%First, we selected all papers referring to the original definition of \hui{} by Feiner~and~Shamash~\cite{feiner1991hybrid} through a Google Scholar search using ``Publish or Perish''~\cite{harzing2007publish}.
%\add{First, we selected all papers referring to seminal works of adjacent research areas (i.e., distributed user interfaces~\cite{elmqvist2011distributed}, complementary interfaces~\cite{zagermann2022complementary}, cross-device interaction~\cite{brudy2019crossdevice}, cross-reality environments~\cite{auda2023scoping}, transitional interfaces~\cite{mayer2024crossing}, augmented displays~\cite{reipschlager2020augmented}) or to the original definition of \hui{} by Feiner~and~Shamash~\cite{feiner1991hybrid}, through a Google Scholar search using ``Publish or Perish''~\cite{harzing2007publish}.}
\add{First, we selected all papers referring to seminal works of adjacent research areas (i.e., \cite{elmqvist2011distributed, zagermann2022complementary, brudy2019crossdevice, auda2023scoping, mayer2024crossing, reipschlager2020augmented}) or to the original definition of \hui{}s~\cite{feiner1991hybrid}, through a Google Scholar search using ``Publish or Perish''~\cite{harzing2007publish}.}
Second, we manually added further publications that \add{were part of the authors' paper collection, discussed among four authors, and eventually identified as potentially relevant.
Some of these handpicked papers did not appear through other search strategies as their individual focus was not the technological setting but the investigated use case (\eg studying window management of virtual displays~\cite{pavanatto2024multiple}).}
Finally, we selected all publications presented at the IEEE ISMAR 2023 Workshop on Hybrid User Interfaces~\cite{hubenschmid2023hybrid}, \add{as contributions to this workshop have explicitly been peer-reviewed in the context of \hui{}s}.
% >>> Results
The three methods resulted in \add{797} publications (see \cref{fig:prisma}).
The cut-off date was \add{December 11th, 2025}.

% ==========================================
% =         Text Graveyard                 =
% ==========================================

% - Since the term hybrid UI is not used consistently in previous work, we conducted a multifaceted literature search using three methods:

% % TODO: maybe as paragraphs, since e.g. the survey results may become too long
% \begin{enumerate}
%     \item
%     We surveyed publications from top publishers related to human-computer interaction: ACM digital library, IEEE xplorer, wiley, and springer.
%     \todo{Search terms... full search terms in appendix..., Potential differences in search engines...}. N=348

%     \item
%     Snowball of original hybrid UI paper N=170
%     % we included papers by searching citations of the \synonym{authoritative} work on hybrid UI~\cite{feiner1991hybrid}.

%     \item
%     We complemented our results with handpicked papers that did not show up in the previous steps but fit the definition of a hybrid UI. N=TODO
% \end{enumerate}

% ``We only considered archival, peer-reviewed papers, \ie journal and conference papers, full and short papers, and book chapters. This approach eliminates workshop summaries, posters, works in progress, and other non-archival publications.''
% \todo{Cut-Off date}

% !TEX root = ../main.tex

% #      ##############################################################
% #      # Figure #####################################################
% #      ##############################################################
\begin{figure}[t!]
    \centering
    \includegraphics[width=.48\textwidth]{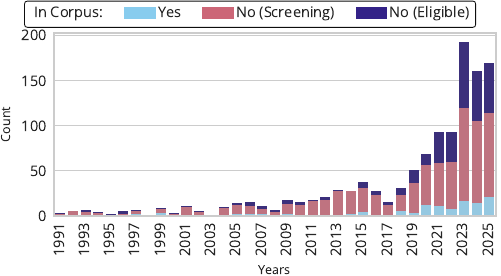}
    \caption{
        %Distribution of records within our corpus over the past three decades.
        Record distribution over the past three decades.
    }
    % \Description{
    %     A stacked bar chart of the distribution of records between 1991 and 2025 and how many records were kept (in sky-blue), or removed from our corpus due to screening (red) or eligibility (purple).
    % }
    \label{fig:corpus}
\end{figure}

% !TEX root = ../main.tex

% ##########################################################
% ######## Sub-Section #####################################
% ##########################################################
\subsection{Selection Process}
\label{sec:prisma:selection}

% >>> Processing
After retrieving the initial corpus of papers (n = \add{1235}), we preprocessed the different output formats and merged them into one table.
% >>> Screening
Three authors screened the reviewed corpus and filtered out duplicates and papers unrelated to MR.
In addition, workshop proposals, complete books, dissertations, or conference proceedings were also removed.

% >>> Eligibility
Subsequently, three authors checked the eligibility of the remaining publications (n = \add{503}).
For that, the following exclusion criteria (based on \cref{sec:rw:aspects}) were defined:
\begin{itemize}
    \item The publication is a duplicate (content-wise) of another (\eg demonstration~\cite{reichherzer2021secondsighta} of system~\cite{reichherzer2021secondsight}).
    \item No prototype or system was presented (\eg discussion of opportunities and challenges~\cite{zaky2023opportunities, zagermann2023challenges, krug2023discussing, reinschluessel2023exploring}).
    \item Only a technological basis is described (\eg frameworks~\cite{hubenschmid2023colibri, sanctorum2019unifying}).
    \item The publication only presents a summary of systems already in our corpus (\eg position papers~\cite{feiner2000environment, feiner1999importance}).
    \item A pure MR system without other \components{} is described (\eg combining AR and VR environments~\cite{cho2023time, bambusek2023how}).
    \item Other device capabilities are only used as input or output modality (\ie{} \components{} are not standalone, \eg attaching input sensors to an AR HWD~\cite{manuri2015preliminary, das2023fingerbutton}).
    \item The publication only presents a collaboration across asymmetric devices (\eg one user on a 2D \component{}, another in an MR environment~\cite{olin2021designing, lunding2022exposar}).
\end{itemize}

% >>> Results
During both selection steps, we split the data set between authors and discussed entries if the decision was unclear.
%Since entries may fulfill multiple exclusion criteria (\eg summary papers also do not present a system) but were rejected once a single criteria was fulfilled, we do not list the amount of papers excluded per criteria.}
This resulted in \add{116} relevant publications within our corpus.

\subsection{Data Extraction\remove{ and Code Book}}
\label{sec:prisma:extraction}

% >>> Intro
\remove{We first created an initial code book to extract data from the remaining 83 papers in our corpus.}

% >>> Codes
Since we see \hui{}s as a sub-category of cross-device interaction, we started with aspects defined by Brudy~et~al.~\cite{brudy2019crossdevice} in their survey.
This was extended with common HCI metadata, such as the contribution type~\cite{wobbrock2016research} or the evaluation strategy~\cite{ledo2018evaluation}.
Furthermore, we recorded each publication's challenges, future work, use case, and devices and terminology used.
% >>> Test coding
To validate the initial set of codes, we \add{randomly} selected 10 papers from our corpus that \add{were coded by} three authors.
This allowed us to (1)~verify if the categorization works, (2)~add missing codes, and (3)~establish a common ground between authors.
% >>> Real coding
With the categories finalized, four authors coded the remaining papers.
%Two authors independently coded every publication.
\add{To ensure the rigor and reliability of the analysis, we employed researcher triangulation: two authors independently coded all publications, and each paper in the final corpus was reviewed by at least two researchers. }
No author assessed the relevance of their own work.
% >>> Grouping / Clean up (Use case, authors term, challenges)
After each paper was coded, the same four authors discussed each publication \add{and clarified conflicts in a coding meeting}.
%and combined the two coding entries (\eg clarifying conflicts).
They further clustered challenges, use cases, and terminology used throughout the corpus.

\subsection{Limitations and Further Considerations}
\label{sec:prisma:limitations}

% >>> Intro
We rigorously designed and conducted our survey.
However, we identified limitations that we present in this section.

% ##########################################################
% ######## SubSub-Section ##################################
% ##########################################################
\ro{Terminology Bias.}
% >>> Intro
\hui{}s can be described by a multitude of adjacent terms (see \cref{sec:rw:adjacent}),
% >>> Problem
making it difficult to create a query that can capture all the systems within this field.
Although we aimed to create the best possible query, this still led to a potentially incomplete set of publications.
% >>> Solution
Therefore, we decided to add other sources (see \cref{sec:prisma:strategy}) to our survey corpus.

% ##########################################################
% ######## SubSub-Section ##################################
% ##########################################################
\ro{Strict Eligibility Criteria.}
% >>> Intro
We searched for a specific type of device combination (\ie MR-enabling devices and conventional 2D displays), leading to strict eligibility criteria (see \cref{sec:rw:aspects}).
% >>> Problem
%However, these can be too restrictive, especially as we also look at the future of \hui{}s in general.
% >>> Solution
To avoid being overly restrictive, we decided to integrate papers that meet our characteristics but leave room for interpretation, especially system papers representing potential directions for future research.
We discuss these edge cases and their implications in \cref{sec:taxonomy:edge_cases}.
% >>> Scale
We acknowledge that our corpus is not exhaustive with regard to possible \hui{} systems caused by the use of said criteria and our intentional focus on \huisubset{}.
Yet, we see our corpus as a representative and substantial set of research on \huisubset{}, showcasing trends that can be generalized to the larger set of papers not captured by our query.
\add{Although no qualitative analysis can guarantee a single definitive taxonomy, we applied an overall methodology that strengthens the validity of our results and ensures that the resulting structure is both rigorous and comprehensive.}

%Following this, we decided to produce a secondary literature corpus out of our initial publication set labeled edge cases.
%These edge cases fulfill a subset of our key aspects, but rule out others.
%While these edge cases fulfill our key aspects, they leave a lot of room for interpretation.
%Commonly, these edge cases included an exclusive usage of complementary components, represented collaboration with asymmetric devices, reproduced reality by VR, or included devices only to benefit from their tangible properties.
%We further discuss and align them in relation to our key aspects in Section \ref{sec:taxonomy:edge_cases} and showcase how they can potentially inform future research with \hui{}s.
%\todo{MS- how do we present those? Where? Are the spatially labeled?}

% !TEX root = ../main.tex

\subsection{Dissemination and Extension}
\label{sec:prisma:extension}

%To allow others to benefit from our survey, 
We have made our literature corpus available using the Indy Survey Tool~\cite{crnovrsanin2023indy} as a GitHub project that hosts an interactive website for users to explore and filter our survey results, as well as to submit other work to our corpus (see \surveyWebsite).
% Please extend if there is more that we should mention. Maybe we could mention the fragmented research landscape again and that we therefore couldn't consider *every* system?

% !TEX root = ../main.tex

% #      ##############################################################
% #      # Figure #####################################################
% #      ##############################################################
% \begin{teaserfigure}
\begin{figure*}[!t]
    \centering
    \includegraphics[width=\textwidth]{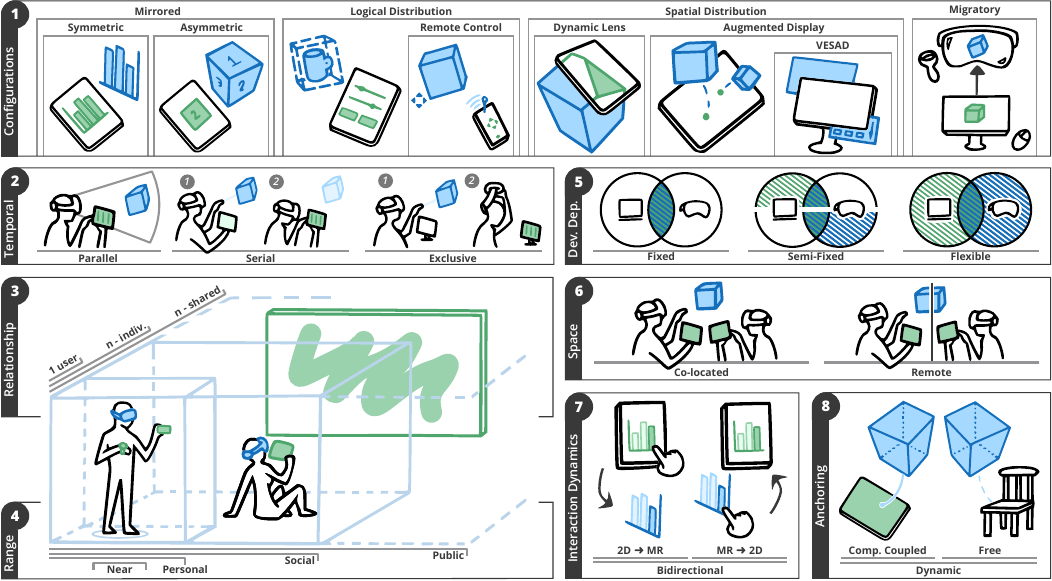}
    \caption{
        We present a taxonomy of key characteristics for mixed reality hybrid user interfaces with eight dimensions: \textit{configuration}, \textit{temporal}, \textit{relationship}, \textit{range}, \textit{device dependency}, \textit{space}, \textit{interaction dynamics}, and \textit{anchoring}.
        \Components{} are highlighted in green for content on a 2D screen and blue for content in a mixed reality environment.
    }
    % \Description{
    %     A taxonomy of augmented hybrid user interfaces with eight characteristics: configuration, temporal, relationship, range, device dependency, space, interaction dynamics, and anchoring.
    %     Each characteristic is visually depicted with an abstract scenario.
    %     Components are highlighted in green for content on a 2D screen and blue for content in a mixed reality environment.
    % }
    \label{fig:dimensions}
\end{figure*}
% \end{teaserfigure}

% !TEX root = ../main.tex

% ##########################################################
% ######## Sub-Section #####################################
% ##########################################################
\subsection{How to Use \add{The Following} Survey}
\label{sec:prisma:how-to-use}

Our survey aims to reveal common patterns and identify shared dimensions within \huisubset{}s.
%With our survey, we aim to reveal common patterns and identify shared dimensions within \huisubset{}s.
Our work can be helpful in three ways:
(1)~By \textbf{classifying past research}, our following \textit{taxonomy of key characteristics} provides a general framework that can be used to describe and understand inherent properties of \huisubset{}s.
%With these dimensions, we can share insights between systems with overlapping dimensions and identify related works within our literature corpus that may not share a common terminology.
(2)~By \textbf{identifying current trends}, our survey shows emerging trends in device combinations, use cases, contribution types, and evaluation strategies.
Readers can understand \textit{how} \huisubset{}s are used, informing the design and evaluation of upcoming \huisubset{}s.
Finally, (3)~our survey can \textbf{inspire future systems}, serving as a roadmap for the next generation of \hui{}s.
%to design the next generation of \hui{}s.

Throughout our reporting (\cref{sec:taxonomy,sec:results}), we denote the amount of literature in our corpus for each characteristic within a \numberboxexample{colored box} mapped on a gradient with the total paper count (n=\add{116}) as the maximum value.
Additionally, we provide a curated set of exemplary papers from our corpus for each characteristic, which can be used as a starting point for further reading.
Since systems can be attributed to multiple characteristics, dimensions, or other categories, the sum of each category does not necessarily match the total number of records within our corpus.

% ==========================================
% =         Text Graveyard                 =
% ==========================================

    %- Based on this set of initial hybrid UI papers, we surveyed papers from relevant publishers that previously published papers on 

% \begin{itemize}
%   \item Creating corpus of relevant publications
%   \item filtering and inclusion criteria
%   \item tagging
%   \item Analysis
%   \item Scale of Survey
% \end{itemize}

% Search Terms:
% \begin{itemize}
%     \item Hybrid User Interface(s)
%     \item Cross-Reality
%     \item Cross-Device Interaction
% \end{itemize}

% Hand-Picked Papers:
% \begin{itemize}
%     \item \cite{feiner1991hybrid} (definition)
% \end{itemize}

% !TEX root = ../main.tex

% ##########################################################
% ######## Sub-Section #####################################
% ##########################################################
\section[A Taxonomy of Mixed Reality Hybrid User Interfaces]{A Taxonomy of Mixed Reality Hybrid User Interfaces}
\label{sec:taxonomy}

Based on our survey, we establish a \textit{taxonomy of key characteristics} for \huisubset{}s (\cref{sec:taxonomy:key_characteristics}) that can characterize existing research and inform new research.
In addition, we describe \textit{emergent trends and opportunities} (\cref{sec:taxonomy:trends})
and highlight \textit{edge cases} (\cref{sec:taxonomy:edge_cases}) of our attributes characterizing \huisubset{}s.

% #####################################################################
% ######## Other Content ##############################################
% #####################################################################
% \input{content/_figures/taxonomy}
% !TEX root = ../main.tex

% ##########################################################
% ######## Sub-Section #####################################
% ##########################################################
\subsection{Taxonomy Dimensions}
\label{sec:taxonomy:key_characteristics}

We adopt six cross-device characteristics of Brudy~et~al.~\cite{brudy2019crossdevice} (\textit{Dimensions 1--6}) and introduce two additional dimensions to better represent the design space of \huisubset{}s.
% We adopt six dimensions of \citeauthor{brudy2019crossdevice}'s cross-device interaction taxonomy~\cite{brudy2019crossdevice}.
To reflect the nuances of \huisubset{}, we extend and reframe each dimension and explain them accordingly.
Although we describe these dimensions as discrete characteristics, they should be seen as continuous spectra, where systems, configurations, and interaction techniques can dynamically span multiple facets.
%Throughout our reporting, we denote the amount of literature in our corpus for each category within a \numberboxexample{colored box}.
%Since systems can be attributed to multiple characteristics, the sum for each category does not necessarily match the total number of records within our corpus, dimension, or other categories.

% ##########################################################
% ######## SubSub-Section ##################################
% ##########################################################
\subsubsection*{\textbf{Dimension 1: Configuration}}

The configuration dimension describes how content and control are distributed across \components{} for a single user.
Although we based our configurations on the cross-device taxonomy (\ie \textit{mirrored}, \textit{logical distribution}, \textit{spatial distribution}, \textit{migratory interfaces}), we identified six configurations unique to \huisubset{}s, namely \textit{asymmetrical mirrored}, \textit{symmetrical mirrored}, \textit{remote control}, \textit{dynamic lens}, \textit{augmented displays}, and \textit{VESADs}.

\begin{description}[leftmargin=0pt]
    \item[Mirrored] configurations duplicate content in different \components{}.
    Given the perceptual differences of components in \huisubset{}s, we distinguish between \textit{symmetric mirror} and \textit{asymmetric mirror} configurations.
    
    \begin{description}
        \item[\descSpacing{} Symmetric Mirror {\normalfont \numberbox{14}}] 
        configurations duplicate content between \components{}, with each \component{} displaying the exact same view of information. %or providing the same functionality.
        This can reduce complexity when interacting with virtual content such as 3D sketching~\cite{arora2018symbiosissketch, drey2020vrsketchin} or immersive analytics~\cite{hubenschmid2021stream, butscher2018clusters}.
        For example, a \hui{} can provide contextual connection based on users' surroundings in MR while providing familiar input on the 2D \component{}. 
        
        \item[\descSpacing{} Asymmetric Mirror {\normalfont \numberbox{15}}]
        configurations show the same view of information but make use of the additional perceptual dimensions offered by MR \components{}:
        %or provide the same functionality on different components.
        % The view of each component is functionally the same but may make use of its extended capabilities:
        For example, this can be used to display a full 3D model or visualization in the MR environment, which is mirrored to an orthographic front view on the 2D \component{}  \cite{szalavari1997personal, iquiapaza2024dear, sereno2022hybrid}.
    \end{description}
    
    \item[Logical Distribution {\normalfont \numberbox{67}}]
    describes configurations where content and control are distributed according to each \component{}'s strength, usually involving a mutually exclusive allocation of responsibilities between \components{}.
    Most systems in our corpus involve some kind of logical distribution, such as offloading text input~\cite{grubert2023text, zhu2020bishare, berns2019myr}), shared and personal space~\cite{rekimoto1999augmented, james2023evaluating}, or general application-control~\cite{satkowski2021insitu, rekimoto1999augmented, nebeling2020mrat}) to 2D \components{}, while the MR environment is used to display content in-situ.
    
    \begin{description}
        \item[\descSpacing{} Remote Control {\normalfont \numberbox{12}}]
        describes a stricter subset of \textit{logical distribution}, where the 2D \component{} provides an alternative (but not exclusive) control over content in the MR environment.
        This can be useful for providing direct interactions when close by and indirect interaction from farther away (e.g., during 3D sketching~\cite{drey2020vrsketchin}) or providing a more ergonomic option~\cite{estalagem2024wearable}.
    \end{description}
    
    \item[Spatial Distribution]
    % configurations provide access to a larger information space through different lenses.
    describes configurations that deliberately spread content across different spatial locations in the continuous real-world space.
    Such a distribution can be achieved through different means, described as \textit{dynamic lenses}, \textit{augmented displays}, or \textit{VESADs}.
    Here, the content is aligned to the 2D \component{} (\ie{} \textit{augmented display}, \textit{VESAD}) or reacts to the position of the 2D \component{} (\ie{} \textit{dynamic lens}).
    
    \begin{description}
        \item[\descSpacing{} Dynamic Lens {\normalfont \numberbox{10}}] 
        allows 2D \components{} to act as a dynamic peephole~\cite{mehra2006navigating} into a larger information space.
        While 2D \components{} provide a constrained view, the MR environment is unrestricted and can use the real environment.
        This requires the 2D \component{} to be spatially aware, enabling, for example, 3D slicing showing a cross-section of a 3D model on the 2D \component{} \cite{luo2021exploring, krug2022clear, szalavari1997personal}.
        
        \item[\descSpacing{} Augmented Displays {\normalfont \numberbox{18}}]
        use the unrestricted visual output of an MR environment to extend a 2D \component{} beyond its potential visual and interaction capabilities, acting as a 3D augmentation that is attached to the 2D \component{}~\cite{langner2021marvis, reipschlager2021personal, chulpongsatorn2023holotouch, dedual2011creating}~.
        This category was initially defined by Reipschläger~et~al.~\cite{reipschlager2020augmented, reipschlager2018debugar} as \textQuote{seamless combination of high resolution touch and pen enabled displays with head-coupled Augmented Reality}.
        
        \item[\descSpacing{} VESADs {\normalfont \numberbox{19}}]
        (\textQuote{Virtually Extended Screen-Aligned Displays}) were initially defined by Normand~and~McGuffin~\cite{normand2018enlarging} as a virtual AR screen \textQuote{that is centered on, and co-planar with, a smartphone}.
        The content is strictly aligned to a 2D \component{} as a seamless display extension~\cite{langner2021marvis, hubenschmid2023smartphone, feiner1991hybrid}, representing a subset of \textit{augmented displays}.
        For example, \textit{VESADs} can be useful for offloading menu elements into MR to save screen real estate on a 2D \component{} \cite{normand2018enlarging, reipschlager2018debugar, brasier2021arenhanced} or provide alternative interaction capabilities~\cite{brasier2021arenhanced}.
        To better represent the diversity of real-world multi-display configurations, we also include configurations that are aligned to 2D \components{} but not necessarily co-planar, such as extending 2D \components{} with angled virtual 2D screens~\cite{reipschlager2019designar, pavanatto2021we, pavanatto2024multiple}.
    \end{description}
    
    \item[Migratory Interfaces {\normalfont \numberbox{22}}]
    enable users to transfer their content or workflow from one device to another.
    Although originally considered asynchronous within the cross-device taxonomy~\cite{brudy2019crossdevice}, \huisubset{}s can also utilize the MR environment to seamlessly transfer content between 2D \components{} \cite{serrano2015gluey} or between a 2D and 3D \component{} \cite{wang2024user, wentzel2024switchspace, aigner2023cardiac, wu2020megereality}.
    In contrast, their asynchronous usage can take advantage of different environments to best support a holistic workflow (\eg by switching between VR and desktop environment~\cite{hubenschmid2022relive, chan2023single}) while still behaving as one unified, continuous system.
\end{description}

% ##########################################################
% ######## SubSub-Section ##################################
% ##########################################################
\subsubsection*{\textbf{Dimension 2: Temporal}}

Prior literature~\cite{brudy2019crossdevice, bornik2006hybrid, hubenschmid2021asynchronous} classifies hybrid and cross-device systems as \textit{synchronous} or \textit{asynchronous}.
Due to the diversity of \components{} within \huisubset{}s, we adopted the suggestion by Bornik~et~al.~\cite{bornik2006hybrid} to further differentiate between \textbf{parallel} and \textbf{serial} usage of \components{}.
Fully asynchronous usage of \components{} was further classified as \textbf{exclusive} for cases where the usage of one \component{} rules out the usage of another (\eg due to spatial or time-related restrictions).

Some papers describe multiple interaction techniques that can be used, for example, in parallel or serial; others only present usages of one distinct aspect.
We report the results accordingly: presenting the number of papers describing mixed and distinct temporal usage first and its subset describing only distinct usage second (\eg{} \numberbox{9} / \numberbox{1}).
%the first number presents papers that can be attributed to a temporal aspect (\ie distinct and mixed usage), whereas the second number represents how many of these records refer to distinct temporal usage (\ie \numberbox{15} / \numberbox{4})}
%presenting the number of papers describing mixed temporal usages first (including single temporal usage) and only distinct temporal usages second (\ie \numberbox{15} / \numberbox{4}).}

%Due to the diversity of components within \hui{}s, we further differentiate between \textbf{parallel}, \textbf{serial}, and \textbf{exclusive} usage of components:

% When referring to hybrid interaction, it is important to differentiate between two approaches: serial and parallel integration. Using serial integration, 2D and 3D methods are used in a sequential order, one after each other. In parallel integration, 2D methods are quasi embedded and used directly to control and adapt the data in the immersive environment. \cite{bornik2006hybrid}

%As multiple temporal codes applied to some papers' descriptions of their, e.g., usage of components, we first report the results including overlaps across the three temporal codes and secondly without overlaps (\ie \numberbox{15} / \numberbox{4}). \todo{MS - rewrite that}

\begin{description}[leftmargin=0pt]
    \item[Parallel {\normalfont \numberbox{94} / \numberbox{65}}] 
    usage indicates that multiple \components{} are used simultaneously, for example, when interacting with one \component{} while observing the output on another \component{} \cite{chulpongsatorn2023holotouch}, transferring content across devices  \cite{hubenschmid2021stream}, or extending a 2D \component{} in the MR environment \cite{langner2021marvis, reipschlager2021personal, hubenschmid2023smartphone}.
    
    \item[Serial {\normalfont \numberbox{45} / \numberbox{9}}] 
    usage indicates that multiple \components{} are used one immediately after another, such as selecting objects of interest in the MR environment and then editing them on a 2D \component{}~\cite{butscher2018clusters}.
    Here, users have immediate access to all \components{} but can focus on only one \component{} at a time to reduce information overload or divide responsibilities between 2D and 3D interaction~\cite{drey2020vrsketchin, feiner1997touring, bornik2006hybrid}.
    
    \item[Exclusive {\normalfont \numberbox{17} / \numberbox{5}}] 
    usage describes asynchronous systems where different \components{} cannot be used simultaneously as part of one workflow but have to be used in sequence.
    This can be useful to bridge the gap between traditional 2D computing environments and VR environments~ \cite{hubenschmid2022relive, chan2023single, schroder2023collaborating}.
\end{description}

\subsubsection*{\textbf{Dimension 3: Relationship}}

The \textit{relationship} category denotes the relation between users within one or multiple systems.
The vast majority of our corpus describes \textbf{single-user} systems~\numberbox{103}:
One user interacts with multiple complementary \components{} (see \cref{fig:dimensions}~\textit{1 user}).
However, we identified several collaborative systems.
We abstracted these into \textbf{multi-user with individual \components{}}~\numberbox{9}:
Each collaborator has their own set of \components{}~\cite{schroder2023collaborating, rekimoto1999augmented} (see \cref{fig:dimensions}~\textit{n-indiv.}); and \textbf{multi-user with shared \component{}}~\numberbox{14}: A display is shared between collaborators as a public space~\cite{mueller2015capture, reipschlager2021personal, krug2022clear, butscher2018clusters} (see \cref{fig:dimensions}~\textit{n-shared}).

% ##########################################################
% ######## SubSub-Section ##################################
% ##########################################################
\subsubsection*{\textbf{Dimension 4: Range}}

We examined the range of interaction across components as defined in the cross-device taxonomy~\cite{brudy2019crossdevice}.
Since the MR \component{} is commonly worn on the user's head (\eg AR HWD) or close to their body (\eg handheld AR), we use this dimension to describe the scale of the 2D \component{} relative to the user.
In addition, the \textit{relationship} can be an indicator of the \textit{range} between \components{} (\eg \textit{multi-user} can indicate \textit{social} or \textit{public} range).
We differentiate between 
    \textbf{near}~\numberbox{8} (\ie \component{} is close to the user's body, \eg smartwatch~\cite{grubert2015multifi, wang2014coordinated, estalagem2024wearable}),
    \textbf{personal}~\numberbox{92} (\ie \component{} is in the personal space, \eg smartphone or tablet~\cite{satkowski2021insitu, hubenschmid2021stream, langner2021marvis, krug2022clear}),
    \textbf{social}~\numberbox{31} (\ie \component{} is in a social space accessible to collaborators, \eg shared display~\cite{butscher2018clusters, mueller2015capture, reipschlager2021personal}),
    and \textbf{public}~\numberbox{2} scale (\ie \component{} can be seen and interacted with by arbitrary bystanders).

% ##########################################################
% ######## SubSub-Section ##################################
% ##########################################################
\subsubsection*{\textbf{Dimension 5: Device Dependency}}

With this dimension, we describe the autonomy of each \component{} within the overall device ecology.
We coded this as \textbf{flexible}, \textbf{semi-fixed}, and \textbf{fixed}:

\begin{description}[leftmargin=0pt]
    \item[Flexible {\normalfont \numberbox{25}}] 
    device dependency indicates that all \components{} provide basic features---a single \component{} could suffice to interact with a system in a meaningful way.
    This can be helpful for workflows that can be easily divided properly into specific subtasks for different environments, such as data analysis workflows~\cite{hubenschmid2022relive}, note taking~\cite{wu2020megereality}, or sketching~\cite{arora2018symbiosissketch}.
    The main value of \hui{}s in this situation comes from the interaction between the components, such as the seamless transition between components.
    
    \item[Semi-Fixed {\normalfont \numberbox{52}}] 
    device dependency represents systems in which one \component{} is completely independent, while others provide supplementary functionality and are thus reliant on the ``main'' component.
    This dynamic has been used primarily to extend the capabilities of existing 2D components, such as increasing the available screen space~\cite{hubenschmid2023smartphone, feiner1991hybrid, langner2021marvis} or offering complementary views on existing content~\cite{berns2019myr, reipschlager2021personal, grubert2015multifi}.
    In contrast, a \textit{semi-fixed} dependency can also be used to provide complementary capabilities to MR environments, such as providing a shared public space~\cite{james2023evaluating} or extracting content from 2D \components{}~\cite{schwajda2023transforming}.
    
    \item[Fixed {\normalfont \numberbox{41}}] 
    device dependency describes systems that can only be used meaningfully with all relevant components present.
    Here, responsibilities are exclusively distributed between components, such as using the 2D \component{} as a haptic surface for touch~\cite{butscher2018clusters, chulpongsatorn2023holotouch, bornik2006interactive}, for contextual interaction within the MR environment~\cite{hubenschmid2021stream, feiner1997touring, satkowski2022investigating} or as a spatially-aware controller~\cite{krug2022clear, luo2021exploring}.
\end{description}

% ##########################################################
% ######## SubSub-Section ##################################
% ##########################################################
\subsubsection*{\textbf{Dimension 6: Space}}

The space dimension describes whether \components{} are \textbf{co-located}~\numberbox{116} (\ie within the same physical space) or \textbf{remote}~\numberbox{0}.
Since it may be difficult to achieve a synchronous \huisubset{} with remote \components{}, all the records we surveyed were exclusively \textit{co-located}.
An asynchronous \huisubset{} with \textit{remote} \components{} may be feasible but would likely forfeit potential benefits gained from any combination of complementary devices (cf.~\cite{brudy2019crossdevice}).
Instead, we see the potential of such remote combinations in collaborative scenarios with \textit{multiple user having individual \components{}} relationships.
Potential remote collaboration scenarios might include individual \huisubset{}s per location and user (\eg a HWD to visualize data in AR and a handheld device for precise input): Here, the AR visualization could be shared and synced, while the handheld devices allow individual manipulations.
However, such scenarios might also require user representations (\eg to create awareness), a communication channel, and a merge policy for conflicting input.
We also discuss opportunities for collaboration using \huisubset{}s in Section~\ref{subsec:collaboration}.

% ##########################################################
% ######## SubSub-Section ##################################
% ##########################################################
\subsubsection*{\textbf{Dimension 7: Interaction Dynamics}}

Since systems in our corpus are intentionally spread across multiple \components{}, the interaction dynamics describe how each \component{} can interact with a system.
Building on the BISHARE~\cite{zhu2020bishare} design space, which classified interaction concepts as either \textit{HWD-centric} or \textit{phone-centric}, we identified three kinds of dynamics:

\begin{description}[leftmargin=0pt]
    \item[Unidirectional (2D-centric) {\normalfont \numberbox{51}}] 
    dynamics, which signifies that input is only possible from the 2D \component{}.
    Examples include touch input~\cite{hubenschmid2021stream, butscher2018clusters, langner2021marvis}, mouse \& keyboard input~\cite{feiner1991hybrid, kim2023perspective, pavanatto2021we}, or using 2D \component{} as spatial controller~\cite{luo2021exploring, krug2022clear} (see \cref{fig:dimensions}~\textit{2D\(\rightarrow\)MR}).

    \item[Unidirectional (MR-centric) {\normalfont \numberbox{7}}] 
    indicates that it is only possible to use input modalities provided by the MR environment, such as controller~\cite{lee2007design, schwajda2023transforming, ilie2004combining} or mid-air gestures~\cite{james2023evaluating} (see \cref{fig:dimensions}~\textit{MR\(\rightarrow\)2D}).

    \item[Bidirectional {\normalfont \numberbox{60}}] 
    dynamics indicate that all \components{} can interact with the system equally, for example, by switching between 2D and 3D sketching~\cite{drey2020vrsketchin}, visualizations~\cite{nebeling2020mrat}, or transferring content~\cite{serrano2015gluey, wu2020megereality}.
\end{description}

% \begin{table}
%     \centering
%     \caption{%
%         \add{%
%         Occurrences of interaction paradigms per \component{} (\textbf{sym}bolic, \textbf{r}eality-\textbf{b}ased-\textbf{i}nteraction, \textbf{dir}ect, \textbf{ind}irect, \textbf{no} paradigm).
%         % The counts are not exclusive, so the sum of a row may exceed its total.
%         Individual records may be attributed to multiple paradigms, so the sum may exceed its total.
%         }%
%     }
%     \begin{tabular}{ccccccc}
%         & \textbf{sym} & \textbf{RBI} & \textbf{dir} & \textbf{ind} & \textbf{no} & \textbf{Total} \\
%         \textbf{AR (OST/VST)}   & \numberbox{67} & \numberbox{30} & \numberbox{45} & \numberbox{15} & 2 & 69 \\
%         \textbf{Handheld AR}    & \numberbox{12} & \numberbox{10} & \numberbox{13} & \numberbox{2} & 1 & 14 \\
%         \textbf{VR}             & \numberbox{18} & \numberbox{11} & \numberbox{13} & \numberbox{9} & 0 & 18 \\
%         \textbf{Spatial AR/VR}  & \numberbox{4} & \numberbox{3} & \numberbox{3} & \numberbox{5} & 4 & 8  \\
%         \textbf{Touch-enabled}  & \numberbox{51} & \numberbox{55} & \numberbox{60} & \numberbox{11} & 2 & 65 \\
%         \textbf{Laptop/Desktop} & \numberbox{1} & \numberbox{34} & \numberbox{1} & \numberbox{34} & 4 & 38 \\
%         \textbf{Projection}     & \numberbox{2} & \numberbox{1} & \numberbox{1} & \numberbox{2} & 7 & 8  \\
%         \textbf{Large Display}  & \numberbox{2} & \numberbox{3} & \numberbox{4} & \numberbox{3} & 4 & 8  \\
%     \end{tabular}
%     \label{tab:interaction-paradigm-counts}
% \end{table}

\add{
    Furthermore, we tagged every \component{} by interaction paradigms that they showcase. For this, we used the non-orthogonal tags: \textbf{symbolic} (WIMP)~\cite{wobbrock2005guess-symbolic}, \textbf{reality-based-interaction} (RBI)~\cite{jacob2008rbi}, \textbf{direct} (pointing at or touching objects to interact), and \textbf{indirect} (interaction target is offset from the location where the input is sensed)~\cite{pfeuffer2015direct-indirect}.
    %(see \autoref{tab:interaction-paradigm-counts}).
    %\autoref{tab:interaction-paradigm-counts} shows the frequency with which these paradigms were employed by each group of \components{}.
    With a few exceptions (\eg fixed study conditions, software toolkits), tracked HWDs and touch-enabled surfaces were tagged with RBI (\numberbox{85} and \numberbox{59}), \eg because of pan/zoom through head/body movement and touch.
    We consider mid-air gestures (\eg grab), pointing, touch, and gaze interaction as direct interaction, while the use of controllers with physical buttons, voice commands, and sign gestures (\eg thumbs-up) are indirect interaction.
    Naturally, laptops and desktop computers almost exclusively relied on WIMP and indirect interaction through mouse/pointers and trackpads (\numberbox{42} in both cases), with a few exceptional peripheral devices.
    This analysis, however, did not reveal unexpected insights about, for example, underutilized \component{} capabilities.
    Regardless, we forward the curious reader to our supplementary material for all details.
}

% ##########################################################
% ######## SubSub-Section ##################################
% ##########################################################
\subsubsection*{\textbf{Dimension 8: Anchoring}}

This dimension describes where the content is placed in the MR environment.
We extended the design space by Reichherzer~et~al.~\cite{reichherzer2021secondsight}, which categorizes content as either world-fixed or device-fixed.
In addition, previous taxonomies have explored anchoring in more detail (\eg semantic and spatial coupling in world-fixed content~\cite{ellenberg2023spatiality}, content layout of device-fixed content~\cite{rashid2012factors, cauchard2011visual}), which we consider out of scope for this work.
Our corpus is split almost equally between three general anchoring techniques:

\begin{description}[leftmargin=0pt]
    \item[Component-coupled {\normalfont \numberbox{32}}] 
    anchoring relates content within the MR environment to the 2D \components{} (\textit{device-fixed}~\cite{reichherzer2021secondsight}).
    To create the illusion of spatial awareness and proper alignment of 2D \components{}, \textit{component-coupled} usually involves spatial calibration of the 2D \component{} for stationary devices or active tracking for mobile platforms.
    \textit{Augmented displays} and \textit{VESADs} always involve a \textit{component-coupled} anchor, while other systems require knowledge about the 2D \component{} for transferring content~\cite{schwajda2023transforming}.

    \item[Free {\normalfont \numberbox{52}}] 
    anchoring describes MR content that is independently placed in the world or attached to real objects in the environment (\textit{world-fixed}~\cite{reichherzer2021secondsight}).
    This method can be used when the 2D \component{} is not directly related to the MR environment, for example, when providing a menu in the 2D \component{}~\cite{drey2020vrsketchin, vock2021idiar} or a simplified but detached view of the MR content~\cite{bornik2006interactive, nebeling2020mrat}.

    \item[Dynamic {\normalfont \numberbox{34}}] 
    anchoring may support both \textit{component-coupled} and \textit{free} anchoring.
    Previous work has explored this approach in terms of transferring content from a 2D \component{} to the MR environment or vice versa~\cite{zhu2020bishare, hubenschmid2021stream, wu2020megereality} or for cutting through 3D models~\cite{krug2022clear, luo2021exploring}.
\end{description}

% !TEX root = ../main.tex

% #      ##############################################################
% #      # Figure #####################################################
% #      ##############################################################
\begin{figure*}[t!]
    \centering
    \includegraphics[width=\textwidth]{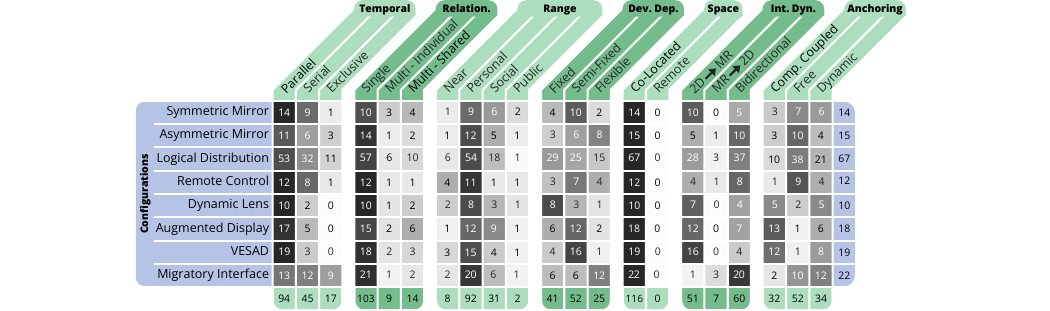}
    \caption{
        The co-occurrences of codes between the \textit{configuration} dimension across the dimensions \textit{temporal}, \textit{relationship}, \textit{range}, \textit{device dependency}, \textit{space}, \textit{interaction dynamics}, \textit{anchoring}.
        Each row is shaded to show the frequency of dimension values for each \textit{configuration}. 
    }
    % \Description{
    %     A table showing the co-occurrences of codes between configurations and other characteristics.
    %     Each row indicates a configuration, while each column depicts a different characteristic from our taxonomy.
    %     Each field is saturated according to the number of relative occurrences for each configuration and dimension, thus highlighting emergent trends.
    % }
    \label{fig:dimensions_table}
\end{figure*}

% !TEX root = ../main.tex

% ##########################################################
% ######## Sub-Section #####################################
% ##########################################################
\subsection{\textbf{Emergent Trends and Opportunities}}
\label{sec:taxonomy:trends}

To discover possible usage patterns of different configurations, we look at the distribution and trends of \textit{dimensions 2--8} across the \textit{configuration} dimension (see \cref{fig:dimensions_table}).
Although existing literature covers a wide range of possibilities, we can identify several trends:
(1)~\textit{Spatial distribution} configurations (including \textit{augmented display} and \textit{VESADs}) have specific requirements, making them unsuitable for \textit{serial} and \textit{exclusive} temporal usage as well as \textit{flexible} dynamics usage.
Their focus on extending a 2D \component{} is indicated by the lack of systems that demonstrate \textit{unidirectional (MR-centric)} interaction dynamics and \textit{free} anchoring.
(2)~\textit{Migratory} configurations can be used regardless of their \textit{temporal} dimension, but have only been explored so far within \textit{single-user} systems.
In this configuration, all \components{} appear to be equally important for interaction, since almost every system in our corpus uses a \textit{bidirectional} interaction dynamics.
(3)~\textit{Unidirectional (MR-centric)} interaction dynamics is only used in configurations that make use of the extended 3D capabilities of the MR \component{} (\ie \textit{asymmetrical mirror}, \textit{logical distribution}, and \textit{migratory interface}).
(4)~Although early work focused mainly on exploring complementary combinations in general \textit{logical distributions}, the increase in hardware sophistication is reflected in an increase in configuration diversity:
For example, \textit{symmetric} configurations only appeared around 2016, while \textit{remote control} configurations appeared around 2020.
Since \hui{}s address an ever-shifting window of opportunity of contemporary hardware capabilities, the choice of possibilities is determined by the available hardware.
%We expect that new hardware will lead to novel combinations, therefore creating new \hui{} \textit{configurations}.
We expect that new hardware will lead to novel combinations, creating new HUI \textit{configurations}.

We can identify several research opportunities by looking at gaps in current usage:
(1)~\huisubset{}s have been exclusively explored in \textit{co-located} spaces.
Although this can be partially attributed to the focus on \textit{single-user} systems, several systems already demonstrate the potential of \huisubset{}s for collaboration.
Similarly, we can see a lack of systems within \textit{public} range.
In both cases, future research could explore the distinct roles of each \component{} in these settings (\eg territoriality, establishing shared and private spaces).
(2)~Only few systems have explored how to interact with 2D \components{} using the MR \component{} (\textit{MR-centric interaction dynamic}).
We see the potential in either \textit{public} scenarios (\eg avoiding hygienic issues) or dynamically enabling remote interaction with 2D \components{}~\cite{horak2018when}:
Here, a \textit{symmetric} configuration may be useful to offer complementary interaction possibilities.
(3)~We attribute the lack of systems with \textit{near} range to current MR hardware limitations.
For example, the limited field of view makes it hard to augment smartwatches.
However, \textit{near} devices could provide a complementary interface to MR HWDs by providing at-a-glance information (cf.~\cite{blascheck2019glanceable}) or touch interaction.
(4)~Since the \textit{logical distribution} configuration represents a substantial amount of records in our survey, we see potential to further differentiate this configuration.
Although such a fine-grained analysis exceeds the scope of this work, future \huisubset{}s might help to reveal additional patterns.

\subsection{Edge Cases}
\label{sec:taxonomy:edge_cases}

We discovered edge cases that were not unambiguously covered by our exclusion criteria (see \cref{sec:rw:aspects}).
We categorize these edge cases into three themes, discuss how they fit into our taxonomy, and sketch potential future directions.

% ##########################################################
% ######## Paragraph #######################################
% ##########################################################
%\paragraph{Collaboration with asymmetric devices}
\ro{Collaboration with asymmetric devices.}
One of our exclusion criteria (see \cref{sec:prisma:selection}) was a collaboration with asymmetric devices (\ie applications where users can interact only with one device in total).
However, such systems can be situated within the cross-device taxonomy~\cite{brudy2019crossdevice} and even be classified as \textQuote{hybrid}, with one user on a tablet and one in AR~\cite{olin2021designing, lunding2022exposar, gottsacker2023hybrid}.
We believe that many of our \textit{key characteristics} are not applicable to such systems.
%, despite their overall hybrid combination.
Although collaborative systems with asymmetric devices might still provide valuable insights for \huisubset{}s, we argue that this would dilute our current focus on cross-device (as opposed to \textQuote{cross-user}) interaction.
Instead, we see them as part of the research field of cross-reality environments (see \cref{sec:rw:adjacent}).

% There is a considerable amount of research on collaboration with asymmetric devices, the combination of which be classified as \textQuote{hybrid} (\eg one user on a tablet, one user in AR~\cite{olin2021designing, lunding2022exposar, gottsacker2023hybrid}).
% %To narrow our scope on the complementarity of components, rather than users, 
% We excluded records where each user can only interact with one device in total, as many of our \textit{key characteristics} are not applicable to such systems, despite its hybrid combination overall:
% While such collaborative systems with asymmetric devices might still provide valuable insights into the design of \hui{}s, we argue that this would dilute our current focus on cross-device (as opposed to \textQuote{cross-user}) interaction.
% Instead, we see such asymmetric collaborations primarily as part of the research field of cross-reality environments (\cref{sec:rw:adjacent}).

% ##########################################################
% ######## Paragraph #######################################
% ##########################################################
%\paragraph{Reproducing reality \& virtual reality}
\ro{Reproducing reality \& virtual reality.}
To avoid limiting our survey to currently available technologies, we also considered systems that simulate or reproduce reality, such as VST HWDs.
Such device combinations are commonly used to overcome the limitations of current OST AR HWDs (\eg increased field of view~\cite{normand2018enlarging}, avoiding different focal planes~\cite{grubert2023text, wieland2024push2ar}), evaluate systems in scenarios that are difficult to reproduce in reality~\cite{jetter2020vr} (\eg supermarkets~\cite{eichhorn2023shopping}), or even simulate hardware capabilities that were not feasible at the time of publication (\eg (transparent) tablets~\cite{krug2022clear, szalavari1997personal}).
Although this can negate some of the benefits of \hui{}s (\eg high-resolution displays of current mobile devices are limited by the clarity of VST HWDs), their conceptual application design contains valuable insights for the complementary use of \components{}, regardless of the technology used.
By extension, we consider combinations that employ a VR environment if they demonstrate complementary use of \components{} according to our \textit{attributes}, such as using a tablet in VR~\cite{surale2019tabletinvr, drey2020vrsketchin}.
In contrast, we excluded papers that did not fulfill our \textit{attributes}, such as ones that do not establish a mutual dependency between the virtual environment and the simulated device~\cite{eichhorn2023shopping}.

% ##########################################################
% ######## Paragraph #######################################
% ##########################################################
%\paragraph{Tangible interaction without visual output}
\ro{Tangible interaction without visual output.}
Mobile devices combined with AR HWDs have potential for tangible interaction techniques that do not necessarily rely on the device's visual output~\cite{stellmacher2024exploring}.
This can be useful to extend the interaction design space.
%(\eg as the mobile device may already be spatially registered).
However, as they do not fulfill our \textit{attributes} (\ie standalone 2D \component{}), we excluded records that do not rely on the mobile device's screen.
%We see such systems primarily within another subset of the broader field of general \hui{}s and the research area of \textit{tangible interaction}~\cite{ishii2008tangible}, which often map digital interactions to dedicated real-world objects.
We see such systems in another subset of \hui{}s, at the intersection of the broader field of \hui{}s and \textit{tangible interaction}~\cite{ishii2008tangible}, which map digital interactions to real-world objects.

\section[How Are Mixed Reality HUIs Used?]{How Are Mixed Reality HUIs Used?}
\label{sec:results}

In this section, we take a closer look at how \huisubset{}s are used throughout our literature corpus.
We highlight findings of our corpus in terms of previously used \textit{terminology} (\cref{sec:results:terminology}), \textit{use cases} (\cref{sec:results:usecases}), \textit{devices and combinations} (\cref{sec:results:device_combinations}), \textit{contribution types} (\cref{sec:results:contribution_types}), and \textit{evaluation strategies} (\cref{sec:results:evaluation_strategies}).
Finally, we provide a \textit{summary of insights} (\cref{sec:results:summary}) \add{and discuss the \textit{potential and pitfalls of \hui{}s} (\cref{sec:results:potential})}.

% #####################################################################
% ######## Other Content ##############################################
% #####################################################################
%% !TEX root = ../main.tex

% ##########################################################
% ######## Sub-Section #####################################
% ##########################################################
\subsection{Terminology}
\label{sec:results:terminology}

We extracted the terminology used by each publication to better understand the terms used to describe \huisubset{}s.
Overall, we found that the author's terminology touched the following areas:
\textbf{Hybrid User Interface}~\numberbox{29}, referring to the term introduced by Feiner~and~Shamash~\cite{feiner1991hybrid};
\textbf{Hybrid \textit{\textlangle{}other\textrangle{}}}~\numberbox{20}, referring to areas such as hybrid computing environment, hybrid display system, or hybrid setup;
\textbf{Cross-Device}~\numberbox{26} and \textbf{Cross-Reality Interaction}~\numberbox{14}, relating to the adjacent research areas;
\textbf{Multi-Device Interaction}~\numberbox{10}, describing a setup related to cross-device interaction;
\textbf{Augmented Displays}~\numberbox{4}, predominately used by Reipschläger~et~al.~\cite{reipschlager2020augmented};
\textbf{Transitional User Interface}~\numberbox{4}, used in relation to cross-reality collaboration or \hui{}s.
In addition, several records used a variety of \textbf{other terms}~\numberbox{13}, such as \textit{compound environment}, \textit{display combination}, or \textit{multimodal interaction}.
Lastly, the rest had \textbf{no explicit terminology}~\numberbox{20}.

% to describe their system.

% !TEX root = ../main.tex

% #      ##############################################################
% #      # Figure #####################################################
% #      ##############################################################
\begin{figure*}[t!]
    \centering
    \includegraphics[width=\textwidth]{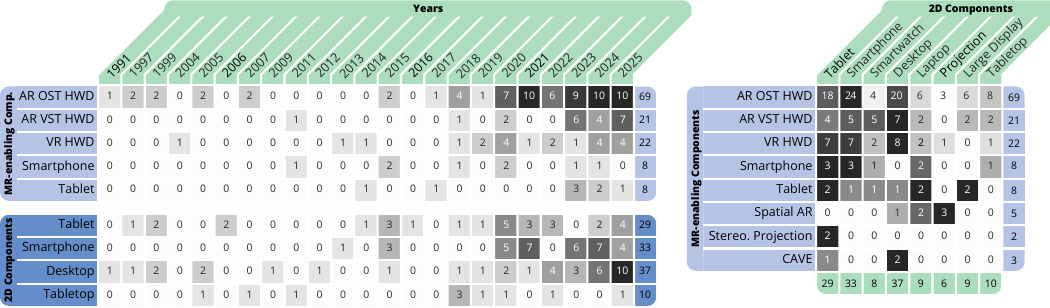}
    \caption{
        (Left)~Distribution of used device types for each year within our literature corpus (years without records were omitted).
        Here, the complete table is shaded based on the maximum value (\ie 10) found in it.
        (Right)~Amount of device combinations within our corpus across a selected set of MR and 2D \components{}.
        Each row is shaded to show the frequency of 2D and MR component combinations.
    }
    % \Description{
    %     Two tables show the number of devices and device combinations.
    %     The left table shows how many devices appeared in each year, with each row depicting one device type and columns depicting each year.
    %     Cells are saturated based on the overall amount of devices.
    %     The right table shows how many device combinations exist in our corpus.
    %     Each row depicts a mixed reality IO component, while each column depicts a 2D IO component.
    %     The cells are saturated again based on the relative amount of combinations for each mixed reality IO component.
    % }
    \label{fig:combinations}
\end{figure*}
% !TEX root = ../main.tex

% ##########################################################
% ######## Sub-Section #####################################
% ##########################################################
\subsection{Use Cases}
\label{sec:results:usecases}

We analyzed usage scenarios as described by the papers in our corpus and further clustered them into common use cases.
%We classified common use cases as described by the authors (\eg prominent example use cases) and clustered these into general application domains and scenarios.
The most prominent use case was \textbf{visual analytics}~\numberbox{33}, ranging from abstract visualization for immersive analytics~\cite{chandler2015immersive} (\eg \cite{hubenschmid2021stream, chulpongsatorn2023holotouch, reipschlager2021personal, satkowski2021insitu, butscher2018clusters, langner2021marvis}), dashboards\cite{vock2021idiar}, user study analysis~ \cite{hubenschmid2022relive, nebeling2020mrat}, and scientific visualizations~\cite{bornik2006hybrid, aigner2023cardiac, luo2021exploring}.
This is unsurprising, since concepts such as multiple-coordinated views are widely established and lend themselves well to configurations such as the \textit{asymmetric mirror}.
This partially overlaps with the \textbf{medical}~\numberbox{7} domain, which uses \hui{}s for 3D examination~\cite{bornik2006hybrid, aigner2023cardiac, luo2021exploring} and surgery~ \cite{ilie2004combining}.
Another popular use case is \textbf{3D modeling}~\numberbox{13}~\cite{surale2019tabletinvr, reipschlager2020augmented, reipschlager2019designar}, including 3D sketching~\cite{arora2018symbiosissketch, drey2020vrsketchin}.
Several records present the area of \textbf{development and authoring}~\numberbox{13}, including development toolkits~\cite{gonzalez2024xdtk, nebeling2020mrat, reichherzer2021secondsight} or programming-related tasks~\cite{reipschlager2018debugar, berns2019myr, wang2014coordinated, lunding2025arthur}.
We attribute several records to general \textbf{productivity}~\numberbox{16} tools, such as window management~\cite{feiner1991hybrid}, extending desktop configurations~\cite{pavanatto2021we, pavanatto2024multiple, kim2023perspective}, file transfer~\cite{serrano2015gluey}, \add{note-taking~\cite{qiu2025marginalia}}, or general user interface improvements~\cite{knierim2021smartphone, brasier2021arenhanced, kane2009bonfire}.
We also found several records in domains such as \textbf{gaming}~\numberbox{5} \cite{tolstoi2015towering, mueller2015capture}, \textbf{entertainment}~\numberbox{6} (\eg music~\cite{elvezio2018hybrid, kohen2020mixr}, television~\cite{kawakita2014augmented, baillard2017multidevice}), \textbf{collaboration}~\numberbox{6} \cite{butz1999enveloping, rekimoto1999augmented, friedl-knirsch2023exploring, schroder2023collaborating}, or \textbf{text entry}~\numberbox{6} \cite{grubert2023text, krug2022clear, bang2023enhancing}.
Lastly, the remainder had \textbf{study-specific}~\numberbox{21} or \textbf{other}~\numberbox{19} use cases.

\subsection{Devices and Combinations}
\label{sec:results:device_combinations}

As publications in our corpus combine multiple \components{} (\ie devices) in a complementary way, we recorded device technologies and their most common combinations (see \cref{fig:combinations}).
For records that did not have specific terminology to describe their hardware (\eg \cite{szalavari1997personal}), we used current terms based on the dimensions of the device.
%In this section we first report on individual types of devices that were used and then continue with describing the most common device combinations.

In terms of MR \components{}, \textbf{AR HWDs}~\numberbox{87} were the most common device technologies, with a large part of systems using \textbf{OST AR HWDs}~\numberbox{69} and others using \textbf{VST AR HWDs}~\numberbox{22}.
For \textbf{handheld AR}~\numberbox{14}, we differentiate between \textbf{handheld AR on smartphones}~\numberbox{8} and \textbf{handheld AR on tablets}~\numberbox{8}.
Other types of AR systems were less common, such as \textbf{projector-based spatial AR}~\numberbox{5}, \textbf{stereoscopic projections}~\numberbox{2}, and \textbf{CAVEs}~\numberbox{3}.
Lastly, since we also included VR environments, our corpus contained systems with \textbf{VR HWDs}~\numberbox{22}.

For 2D \components{}, we observed a similar spread between available device technologies:
Several systems used mobile platforms~\numberbox{66} such as \textbf{smartwatches}~\numberbox{8}, \textbf{smartphones}~\numberbox{33}, \textbf{tablets}~\numberbox{29}, and \textbf{laptops}~\numberbox{9}.
Stationary \components{}~\numberbox{59} were almost as common, including \textbf{desktops}~\numberbox{37}, \textbf{projectors}~\numberbox{6}, \textbf{large displays}~\numberbox{9} (\eg wall-displays), and \textbf{tabletops}~\numberbox{10}.

Looking at the device combinations (see \cref{fig:combinations} left) over the years reveals an increase in device variation as more new form factors become available.
This proliferation resulted in a variety of device combinations (see \cref{fig:combinations} right), with a focus on HWDs, although other MR devices were also used.
Most often, 2D \components{} were represented by handheld devices (\eg smartphones), or desktop \components{}.
The usage of \textbf{AR OST HWDs} was preferred for the described 2D \components{}.
\subsection{Contribution Types}
\label{sec:results:contribution_types}

We classified papers in our corpus following the definition of Wobbrock~and~Kientz~\cite{wobbrock2016research}.
%(see \cref{fig:contributions}).
As some papers provide more than one contribution (\eg an artifact that was used as an apparatus in an empirical user study), we distinguish between primary and secondary contributions and report the results accordingly (\ie \numberbox{15} / \numberbox{4}). 

\begin{description}[leftmargin=0pt]
    \item[Artifact {\normalfont \numberbox{73} / \numberbox{11}}]
    contributions manifest new knowledge in a design-driven approach creating new systems, tools, and techniques.
    See Section~\ref{sec:results:device_combinations} for further descriptions of, for example, device combinations used to create artifacts.
    
    \item[Empirical {\normalfont \numberbox{36} / \numberbox{19}}]
    contributions provide new knowledge in an evaluation-driven approach based on user studies.
    See also Section~\ref{sec:results:evaluation_strategies} for descriptions of evaluation strategies.
    
    \item[Theory {\normalfont \numberbox{6} / \numberbox{6}}]
    contributions improve existing concepts, creating frameworks.
    We consider thorough descriptions of design spaces as theoretical contributions.
    
    \item[Method {\normalfont \numberbox{1} / \numberbox{1}}]
    contributions create new knowledge that informs how researchers carry out their work.
    
    \item[\add{Dataset} {\normalfont \numberbox{1} / \numberbox{0}}]
    \add{contributions offer a curated corpus for a specific topic to help design future \huisubset{}s and layouts.}
\end{description}

Two papers in our corpus~\cite{langner2021marvis, zhu2020bishare} are each categorized with a single primary contribution and two equivalent secondary contributions:
Both present a design space as their main contribution (theory), followed by a system (artifact) used as the apparatus of a user study (empirical).

No papers were classified as surveys or opinion contributions.
%(see \cref{fig:contributions}).
This is not surprising, as we focused on actual systems, which resulted in most artifact and empirical contributions.
Similarly, some of the papers that could fulfill the criteria of other types of contribution were not considered, as they typically did not meet the eligibility criteria of our survey. 
Five papers contributed a theoretical contribution (\ie a design space~\cite{langner2021marvis, zhu2020bishare}), showing that design spaces can be a way of exploring this nascent research area without necessarily implementing a system or study apparatus.

%Limitation: our search was focused on systems, therefore mostly artifact and empirical contributions -> \textQuote{Theory or framework paper types were not considered as they often did not meet the eligibility criteria for our survey}.
%- Still, X\% methodological contributions -> Design Spaces can be a good way to better explore this nascent space, without exploding complexity / scope of prototype

% !TEX root = ../main.tex

% ##########################################################
% ######## Sub-Section #####################################
% ##########################################################
\subsection{Evaluation Strategies}
\label{sec:results:evaluation_strategies}

For papers that provided a primary or secondary empirical contribution, we coded their evaluation strategies~\cite{ledo2018evaluation}.
%(see \cref{fig:contributions}).
Most papers featured one empirical contribution; others combined multiple user studies with different strategies.

The most common evaluation strategies within our corpus were empirical user studies that examine how users interact with a system~\numberbox{73}.
In their cross-device taxonomy, Brudy~et~al.~\cite{brudy2019crossdevice} further split this evaluation strategy into \textit{qualitative and quantitative usage} and \textit{informative (observational and elicitation)} evaluations:
While \textbf{usage}~\numberbox{59} focuses on the usability and usefulness of the system and how it is appropriated~\cite{brudy2019crossdevice}, \textbf{informative}~\numberbox{23} evaluations involve studies that precede and inform the development of a system, involving users in the design process~\cite{brudy2019crossdevice}.
\textbf{Demonstrations}~\numberbox{21} are used to describe how systems are employed in an actual use case scenario but do not necessarily involve a real system implementation.
In contrast, \textbf{technical performance}~\numberbox{5} evaluations (cf. \textit{technical evaluation}~\cite{brudy2019crossdevice}) focus on benchmarking an implemented system in terms of its technical capabilities.
Lastly, several systems did not include any kind of evaluation~\numberbox{21}.
Similarly, we did not observe any kind of \textbf{heuristics}~\numberbox{0} evaluation (\ie using guidelines to analyze usability), which could be attributed to the lack of appropriate guidelines for \huisubset{}s.

% !TEX root = ../main.tex

% #      ##############################################################
% #      # Figure #####################################################
% #      ##############################################################
% \begin{figure}[t!]
%     \centering
%     % \includegraphics[width=\columnwidth]{figures/year X contribution X evaluation.pdf}
%     \includegraphics[width=.48\textwidth]{figures/year_X_contribution_X_evaluation.pdf}
%     \caption{
%         Distribution of primary contribution types and evaluation strategies over time within our corpus (years without records were omitted).        
%         Here, each cell within a row (\ie year) is shaded based on how many papers presented this contribution or evaluation type in the respective year, visualizing the relative development over time (\ie in a column).
%     }
%     % \Description{
%     %     A table showing the amount of contribution types and evaluation strategies between 1991 and 2024 within our corpus.
%     %     Years are depicted as rows, while columns depict the contribution types and evaluation strategies.
%     %     Each field is saturated according to the number of relative occurrences for each contribution type and evaluation strategy.
%     % }
%     \label{fig:contributions}
% \end{figure}

% !TEX root = ../main.tex

% ##########################################################
% ######## Sub-Section #####################################
% ##########################################################
\subsection{Summary of Insights}
\label{sec:results:summary}

We present four main insights from our literature survey.

% ##########################################################
% ######## Paragraph #######################################
% ##########################################################
% \paragraph*{Term Fragmentation}
\ro{Term Fragmentation.}
%
%Our corpus shows that terminology in prior research is mostly evenly split between \textit{hybrid user interfaces}~\numberbox{29} and \textit{cross-device interaction}~\numberbox{65}, while the emergent area of \textit{cross-reality}~\numberbox{14} is gaining traction.
\add{Our corpus shows that terminology in prior research mostly favors the broader term \textit{cross-device interaction}~\numberbox{65} over \textit{hybrid user interfaces}~\numberbox{29}}, while the emergent area of \textit{cross-reality}~\numberbox{14} is gaining traction.
A large number of eligible records in this survey were found in other research areas, barely mentioning such \textit{hybrid} device combinations and thus making it difficult to search for and identify relevant prior work.
This fragmentation is also reflected in their specificity, ranging from rather broad (\eg \textit{cross-device interaction}~\numberbox{26}, \textit{multi-device interaction}~\numberbox{10}) to very narrow (\eg \textit{hybrid \textlangle{}other\textrangle{}}~\numberbox{20}) descriptions.
While broad terms encapsulate a breadth of unrelated systems, overly specific denominators may lead to fragmentation and could impede the understanding of the field.

A similar fragmentation can be seen in the use of terminology to describe design dimensions.
Although the cross-device taxonomy~\cite{brudy2019crossdevice} provides an appropriate framework, we adapted many dimensions to better capture the unique design dimensions of \huisubset{}s.
Many useful terms that apply to \huisubset{}s are hidden within artifact contributions (\eg anchoring~\cite{reichherzer2021secondsight}, dependency~\cite{zhu2020bishare}, configurations such as augmented display~\cite{reipschlager2020augmented, reipschlager2018debugar}, and VESADs~\cite{normand2018enlarging}).

% ##########################################################
% ######## Paragraph #######################################
% ##########################################################
% \paragraph*{Parallel usage is predominant}
\ro{Parallel usage is predominant.}
We found that \textit{parallel} usage of multiple \components{} is by far the most predominant design choice.
We think the reasons are twofold.
(1)~\textit{Parallel} usage offers greater design possibilities.
In contrast, \textit{exclusive} (and to some extent \textit{serial}) usage considers one \component{} at a time, thus limiting the design potential.
(2)~As the time between using different \components{} \textit{exclusively} increases, it becomes harder to see the whole system as one coherent interface---and harder to classify.
Systems with \textit{exclusive} usage might be better described as \textit{cross-reality}, which is concerned with the general usage of multiple systems in different \textQuote{realities}.

% ##########################################################
% ######## Paragraph #######################################
% ##########################################################
% \paragraph*{Optical see-through is the prevalent hardware choice, despite its drawbacks}
\ro{Optical see-through is the prevalent hardware choice, despite its drawbacks.}
AR HWDs~\numberbox{87} were the most common MR \component{}.
Although current VST HWDs~\numberbox{22} offer a wider field of view for digital content~\cite{normand2018enlarging}, the use of OST HWDs~\numberbox{69} was much more prevalent.
We attribute this to
(1)~the unrestricted real-world field of view of OST HWDs greatly facilitating interaction with 2D \components{};
(2)~using OST HWDs further emphasizes the complementary nature of \hui{}s, as the addition of a secondary \component{} offsets the drawbacks of the AR HWD;
and (3)~VST HWDs having only recently matured enough to be used in conjunction with other devices (\eg in terms of text legibility due to limited pass-through resolution).
%However, recent improvements in VST hardware (\eg significantly better pass-through resolution and latency) may shift this trend towards VST HWDs.
%However, recent commercial VST HWDs have improved significantly (\eg Apple Vision Pro) while commercial OST HWDs have stagnated, which may cause a shift towards 
We also observed a steady increase in empirical contributions over the past years, indicating that the hardware is now mature enough to conduct studies that are not confounded by hardware restrictions.

% ##########################################################
% ######## Paragraph #######################################
% ##########################################################
% \paragraph*{Lack of collaborative systems}
\ro{Lack of collaborative systems.}
Our corpus shows a distinct lack of multi-user systems~\numberbox{21}, especially within public spaces~\numberbox{2}.
Although there is a great deal of work in the field of computer-supported cooperative work~(CSCW) for collaboration across asymmetric devices---which we intentionally excluded from our survey---we see great potential in the use of \huisubset{}s in collaborative scenarios.

% !TEX root = ../main.tex

\subsection{\add{Potential and Pitfalls of \huisubset{}s}}
\label{sec:results:potential}

\add{
\huisubset{}s are an intentional combination of off-the-shelf devices that would otherwise be used on their own.
This is one of their strengths:
Instead of requiring bespoke hardware tailored to a narrow range of tasks (\eg tangibles), \huisubset{}s leverage the user's existing device ecology.
%, with each device already being capable of solving the task independently---even if, at times, poorly.

However, this can also have several pitfalls:
Users' preferences for modalities might not necessarily reflect the design of the \huisubset{}, as they might, for example, favor efficiency over effectiveness (cf. \cite{zagermann2017memory}).
In addition, as off-the-shelf devices are designed for general-purpose use, they inevitably involve trade-offs.
For example, a smartphone's screen may remain mostly unused inside MR environments, as users focus their visual attention on the content of the MR HWD.
%Would users not be better served by employing specialized technologies, such as tangibles or bespoke input devices?
Should we favor specialized technologies, such as bespoke input devices, over general-purpose ones?

While specialized technologies may offer better performance for their intended use case, they lack the versatility to be used outside of their designated environments, leaving users with a plethora of single-purpose devices.
In contrast, \huisubset{}s leverage devices that are already within the user's ecology, connecting not only hardware capabilities but also software ecosystems.
This could increase the adoption of immersive technologies, as they can be integrated to enhance current workflows and practices, rather than supersede them.
Even outside of the context of \huisubset{}s, these devices remain valuable, as smartphones and MR HWDs, for example, can still be used on their own.

\huisubset{}s thus represent a pragmatic design solution that leverages the complementarity of currently available device technologies within existing device ecologies.
Even as hardware limitations become less relevant due to increasingly sophisticated MR HWDs, \huisubset{}s can inform the design of future interaction techniques and configurations.
For example, recent advances in MR HWDs may render migratory configurations obsolete, as users can simply simulate a desktop in MR.
Yet, using \huisubset{}s, we can already investigate potential techniques and devices (\eg hybrid input devices~\cite{hubenschmid2025spatialmouse, hubenschmid2025revisiting}) that transcend the use of \huisubset{}.
}

\section{Key Challenges and Research Agenda}
\label{sec:discussion}

Our thorough literature survey and our own experiences allow us to discuss key challenges and research opportunities.
\subsection{Transitioning Between \Components{}}

\hui{}s draw their strength from the combination of complementary \components{}.
While this can be beneficial, it also introduces several challenges concerning \textit{visual attention switching}, \textit{transferring content between \components{}}, and \textit{perceptual content synchronization}.

% --------------------------------------
% \paragraph*{Visual attention switching}
\ro{Visual attention switching.}
Introducing multiple visual output components divides the user's attention among each component, increasing mental load~\cite{rashid2012cost}.
Prior work has investigated such visual attention switches in multi-display environments~\cite{rashid2012factors} and proposed metrics to potentially reduce strain.
Although \huisubset{}s can benefit from these insights, their design space is much broader in terms of display layout, enabling display layouts that are unbound by reality.
Still, \add{factors such as distance~\cite{bang2025areading} or} current hardware limitations have to be considered, as several works~\cite{eiberger2019effects, grubert2015multifi, normand2018enlarging, wieland2024push2ar} opt for VST AR HWDs to avoid differences in focal planes, which can further increase mental load~\cite{eiberger2019effects}.
Overall, more research is required to investigate potential effects and solutions for different display layouts~\cite{knierim2023universal}.

% --------------------------------------
% \paragraph*{Transferring content between \components{}}
\ro{Transferring content between \components{}.}
\huisubset{}s mix 2D and 3D content to make use of each \component{}'s strengths.
While the distribution is often static (\eg determined by the designer of the system), several systems have demonstrated the utility of transferring content from a 2D \component{} to the MR environment~\cite{schwajda2023transforming, wu2020megereality, wang2024user, aigner2023cardiac}.
Specifically, prior work~\cite{schwajda2023transforming} highlights the importance of animations and linking when ``pulling'' content from a 2D \component{} into an MR environment (cf. showing Bézier curves to selected MR content~\cite{hubenschmid2021stream, vock2021idiar}).
Furthermore, recent research in the area of immersive analytics investigated visualization transformations between 2D and 3D~\cite{lee2022design, lee2023deimos}.
Several transition metaphors have been \add{explored~\cite{wu2020megereality,zhao2025spatialtouch} and compared~\cite{rau2025traversing,cools2025comparison}} in prior work, but further investigations are necessary to standardize \remove{and evaluate} these techniques.

% --------------------------------------
% \paragraph*{Perceptual synchronization}
\ro{Perceptual synchronization.}
Another challenge is found within the mapping and synchronization of content and interaction across components.
While prior research has shown that an asymmetry of interaction (\eg decoupling spatial interaction~\cite{sereno2022hybrid}) or information (\eg showing simplified 2D views of 3D content \cite{hollerer1999exploring, iquiapaza2024dear}) can be beneficial, this may come at an increased cost of user perception:
For example, how does a 2D interaction affect its 3D equivalent, and how can we communicate this asymmetry?

Aside from these conceptual challenges in establishing a consistent mental model, we also have to consider technological challenges.
In this context, prior work found that cross-device interaction techniques can be highly sensitive to network latency~\cite{liu2023defining}.
This sensitivity may be even more pronounced in \huisubset{}s, especially if they appear as one conceptual device (\eg the virtual screen in a \textit{VESAD} configuration may lag behind the real screen).
Here, more research is necessary to investigate how inevitable technological factors (\eg latency) can confound findings.

% --------------------------------------
% --------------------------------------
% --------------------------------------
\subsection{Unchaining Device Capabilities}

The past decade has seen a surge in both device variety (\eg smartwatches, smartrings) and device capabilities (\eg inside-out spatial tracking), which can also be seen in the increase in \component{} combinations.
Yet, 2D \components{} such as smartphones are restricted by their form factor, favoring device ergonomics and portability but limiting their potential output capabilities.
By combining them with MR \components{} (\eg AR HWDs), we can ``unchain'' their capabilities, enabling entirely new interaction possibilities (\eg offloading menu items~\cite{reipschlager2019designar, normand2018enlarging, brasier2021arenhanced}), thereby inching closer to a form of ``universal interaction''~\cite{knierim2023universal} and ultimately, Weiser's vision of ubiquitous computing~\cite{weiser1999computer}.
While this can further expand the design space, we also recognize the need to establish guidelines to better understand the trade-offs and potential of each device (cf. \cite{wang2023serious}).

% --------------------------------------
% --------------------------------------
% --------------------------------------
\subsection{Evaluation, Assessments, and Toolkits}

Brudy~et~al.~\cite{brudy2019crossdevice} described that research on cross-device interaction can be considered as a \textQuote{constructive problem}~\cite{oulasvirta2016hci}.
This is reflected in our corpus:
The majority of papers focus on creating new artifacts.
%---systems, tools, and techniques.
These papers \textQuote{push the boundaries of interaction possibilities}~\cite{brudy2019crossdevice}---a common theme for research in HCI that is often \textQuote{much better at proposing new technologies than at validating them}~\cite{hornbaek2011whys}, especially when novelty is expected as a key contribution\footnote{\url{https://sigbed.org/2022/08/22/the-toxic-culture-of-rejection-in-computer-science/} last accessed on \lastaccess}.
This further fragments research on \hui{}s, as artifact and empirical contributions tend to drift apart~\cite{brudy2019crossdevice}.
%While some research addressed the call of Brudy~et~al.~\cite{brudy2019crossdevice} for a frame of reference to compare cross-device interaction techniques~\cite{zagermann2020otherhand}, this is still missing for research on \huisubset{}s.
While some research addressed the call for a frame of reference~\cite{brudy2019crossdevice} to compare cross-device interaction techniques~\cite{zagermann2020otherhand}, this is still missing for research on \huisubset{}s.
One possibility to create such frames of reference and systematically study (and later compare) research within this space is to consider \textit{experiments as design-informing activities}~\cite{oulasvirta2022counterfactual}.
Here, different design alternatives (\eg meaningful combinations of input and output components) can serve as independent variables for different use cases.
This allows us to study their effect on general (\eg~time, error) and use case-specific dependent variables, such as utilization of devices~\cite{schroder2023model}---focusing on the effect, influence, and utility of each combination, beyond a technical perspective~\cite{brudy2019crossdevice}.

\hui{}s differ from single-device user interfaces by involving multiple \components{}, making assessments more complex.
For example, cognitive workload (a typical metric in HCI user studies~\cite{kosch2023survey}) can be measured post-hoc using subjective questionnaires like NASA TLX~\cite{hart1988development, hart2006nasatask}, but real-time, objective methods such as eye tracking are preferable~\cite{kosch2023survey}.
As \huisubset{}s often include an HWD, continuous assessment of eye movements via built-in sensors is achievable.
However, switching between \components{} complicates data collection, as some \components{} might not support eye tracking or may have varying data quality, requiring data fusion or repeated calibration.
This issue becomes more complex with multiple users frequently switching between \components{}.
Although subjective questionnaires simplify user studies, they reduce data accuracy, as participants might forget specific details, especially in highly dynamic \huisubset{}s.
Objective, real-time assessments are critical for understanding \huisubset{}s, but careful study designs are necessary to manage complexity in conducting and analyzing them.

To address this complexity, research has suggested a variety of toolkits that can support this process for MR user studies~\cite{hubenschmid2022relive, reipschlager2022avatar, buschel2021miria, nebeling2020mrat}.
Yet, these toolkits also hold the potential to support the design, conduction, and analysis procedures of user studies involving \huisubset{}s.

% --------------------------------------
% --------------------------------------
% --------------------------------------
\subsection{Authoring Mixed Reality Hybrid User Interfaces}

Creating \huisubset{}s can be challenging:
Designers require evidence-based guidelines that focus on the integration of multiple devices~\cite{daeijavad2023hybrid}, such as determining the optimal ergonomic content distribution~\cite{zaky2023opportunities, belo2021xrgonomics}.
Yet, developers also need to deal with implementations on different platforms.
Although web technologies (\eg WebXR) can standardize development, their support on commercial HWDs is uncertain~\cite{butcher2023don}.
In addition, toolkits (\eg for integration of multiple devices~\cite{gonzalez2024xdtk}, \component{} configuration~\cite{sandor2005immersive}, synchronization~\cite{hubenschmid2023colibri}) can help focus on the implementation at hand, rather than worrying about implementation details.
Here, a common grammar (cf. VEGA~\cite{satyanarayan2016reactive, satyanarayan2017vegalite}) could allow content to \textit{responsively} adapt to each device.

% --------------------------------------
% --------------------------------------
% --------------------------------------
\subsection{Exploring Holistic Real Life Applicability}

The increase in artifacts and evaluations within our corpus indicates that the nascent space represented by our corpus is slowly maturing.
Research prototypes and their interaction techniques have so far been investigated in isolation, yet \textQuote{it is clear that [these] facets should not be discussed in isolation; instead, they are highly interconnected and affect each other}~\cite{krug2023discussing}.
Given the wide applicability of existing (see \cref{sec:results:usecases}) and future systems within our \hui{} subset (\eg robotics~\cite{lunding2024proposing}, medical domain~\cite{reinschluessel2023exploring, krug2023discussing, smith2023integrating}, explainable artificial intelligence~\cite{mendez2023how}), we need to consider their role within the holistic context of their work environments and unveil their unique challenges and opportunities.
Furthermore, consolidating common interaction techniques into a library of design patterns can help researchers and practitioners alike in the design and implementation of holistic \huisubset{}s.

% --------------------------------------
% --------------------------------------
% --------------------------------------
\subsection{Collaboration}

\label{subsec:collaboration}
The majority of the papers in our corpus focus on \textit{single-user} systems.
Yet, some papers indicated a collaborative setting, either with a shared \component{} (\eg a shared output) or using individual \components{}.

Regarding co-located collaboration, the choice whether there is a shared \component{} or individual \components{} can directly influence the type of collaboration:
For example, Butscher~et~al.~\cite{butscher2018clusters} used a large interactive tabletop display as a shared input device, combined with an AR HWD for each user.
Limited by the touch input of the shared device, only one user is able to manipulate the content, thus enforcing a closely coupled collaboration~\cite{gutwin1998design, tang2006collaborative}.
In contrast, providing users with individual \components{} enables loosely coupled, parallel activities~\cite{schroder2023collaborating}.
Similarly, asymmetric device setups might benefit from individual device affordances~\cite{friedl-knirsch2023exploring} but also lead to an asymmetry of roles~\cite{zagermann2023challenges}.

Lastly, no papers in our corpus feature remote collaboration.
In such cases, techniques known from MR remote collaboration could be applied:
Virtual avatars can represent the remote person to create awareness~\cite{gronbaek2024blended, gronbaek2023transitional} while methods to align workspaces can allow for deictic referencing~\cite{fink2022relocations}.
The use of diverse \components{} might require user representations per \component{}, and highly dynamic environments might benefit from transitional user representations~\cite{zagermann2023challenges}.

\section{The Future of Hybrid User Interfaces}
\label{sec:outlook}

Our literature survey has shown the past and present of \huisubset{}s, allowing us to infer trends for the larger domain of \hui{}s.
However, one question still lacks clarity:

\ro{So, what is a \hui{} now?}
The term ``hybrid'' in \hui{} introduces ambiguity, as it could describe any combination of multiple interfaces and concepts.
This ambiguity and divergent facilitation of the term is further demonstrated by contrasting the original definition of \hui{}s---referring to \textQuote{heterogeneous display and interaction}~\cite{feiner1991hybrid} or \textQuote{different interface}~\cite{feiner1994redefining} technologies---with its most prevalent use today (\ie combining 2D and MR components).
This led to our decision to not sharply define \hui{}s and \huisubset{}s because of the 
    (1)~historical developments and usages of the term, 
    its (2)~relations and overlaps to adjacent research areas, 
    (3)~the abundance of possibilities for combinations as shown in our taxonomy, 
    and (4)~potential edge cases.
However, we opted to modernize Feiner and Shamash's definition of \hui{}s~\cite{feiner1991hybrid} (see \cref{sec:rw:definition}) while specifically focusing on the initial device combinations of 2D and MR \components{}.

Through our work on this survey and our own experience, we can say that the term \hui{} (and with that \huisubset{}) is more \replace{a kin}{akin} to a \textbf{fuzzy concept}---\textQuote{a collection of objects [that do] not have sharp, clear-cut boundaries}~\cite{belohlavek2011concepts}.
Thereby, \hui{}s can be characterized with certain attributes as described in \cref{sec:rw:aspects}. %, defining \textQuote{the membership [as] a matter of degree}~\cite{belohlavek2011concepts} compared to an arbitrary fixed inclusion or exclusion based on currently available device technologies.
In the end, actual manifestations may vary---depending on the discussed ever-shifting window of opportunities---where \hui{} boundaries remain adaptable to emerging interaction paradigms, technological affordances, and contextual demands.
However, if the term is adaptable in such a way, we must ask:
% While it is common to look at the past to understand the present and predict the future, we must ask:

%\vspace{-.6em}
\ro{Will there be \hui{}s in the future?}
%
%While it is a common theme to look at the past to understand the present and potentially predict parts of the future, one could ask a rather provocative question: \textQuote{Will there be \hui{}s in the future?}
%From a positive perspective, 
On the bright side, \hui{}s combine complementary device technologies, taking advantage of the strengths of each technology, creating user interfaces where \textQuote{the real power \elide{} comes not from any one of these devices; it emerges from the interaction of all of them}~\cite{weiser1999computer}.
This combination can create flexible and modular systems in which the user can freely decide when to employ which type of feature (\ie \component{}).

However, from a rather critical perspective, \hui{}s could be considered as a compensation mechanism for the deficiencies of individual \components{}, such as the limited precision of gesture recognition in AR environments.
By including additional technologies to address the shortcomings of any one input method, these systems can become too complex to design, evaluate, and eventually use.
With advances in technology, the need for \hui{}s composed of multiple \components{} may diminish:
An ultimate interface might not require the combination of multiple \components{} to compensate for individual weaknesses.
Instead, at some point in the future, there will be self-contained ``devices'' that streamline interaction into a single cohesive system.
This leads to the question:

%\vspace{-.6em}
\ro{Is there an expiration date for \hui{}s?}
We believe so, but cannot say when.
Over 30 years of research on \hui{}s and related research streams have shown a maturing of concepts and usage of technology.
% >>> What has changed?
Although some of the pioneering works are often technology-driven, proving the feasibility of hardware-heavy combinations, newer works are more mindful of the usage of technology.
At the same time, we observed an increase in various evaluation strategies over time.
% (see \cref{fig:contributions}).
% >>> Why though
Both can be attributed to advances in technology (\eg advanced built-in tracking capabilities)---minimizing the need for additional technical augmentation while focusing more on the user of such systems.
This shifts the focus from purely technical work towards studying the individual benefits of various design alternatives (\eg different variants of \hui{}s) and emphasizes the value and utility of individual \components{} and the user experience of \hui{}s. 
% >>> To what did that lead
This maturing process led to current off-the-shelf devices (\eg Apple Vision Pro\footnote{\url{https://apple.com/apple-vision-pro/}, last accessed on \lastaccess}) and applications (\eg Immersed\footnote{\url{https://immersed.com}, last accessed on \lastaccess}), which already bring \hui{}s to the consumer market.
With the maturation of concepts, an increase in empirical contributions, the advent of consumer-oriented \hui{}s, and the development of research areas such as artificial intelligence or even brain-computer interfaces, we can take a step back and look at the bigger picture with a last question:

% The over 30 years of research on \hui{}s and related research streams have shown a maturing of concepts and usage of technology.
% While some of the pioneering works are often technology-driven, proving the feasibility of hardware-heavy combinations, newer works are more mindful of the usage of technology.
% This can be attributed to advances in technology (\eg advanced built-in tracking capabilities) -- minimizing the need for additional technical augmentation of \hui{} environments.
% However, our literature survey has shown an increase in various evaluation strategies for \hui{}s over time (see \cref{fig:contributions}).
% With this, the focus shifted from purely technical work towards studying the individual benefits of various design alternatives (\eg different variants of \hui{}s) to emphasize the value and the utility of individual components and the user experience of \hui{}s. 
% Similarly, current off-the-shelf devices (\eg Apple Vision Pro\footnote{\url{https://www.apple.com/apple-vision-pro/}, last accessed on \lastaccess}) and applications (\eg Immersed\footnote{\url{https://immersed.com}, last accessed on \lastaccess}) already bring \hui{}s to the consumer market: Enabling a larger display space by \textit{migrating} the visual output from a 2D display to AR.

%\vspace{-.6em}
\ro{What does the future of \hui{}s look like?}
%
% >>> What is now
With our attributes, we identified a classification of past and current \huisubset{}s.
The interplay of \components{} is fundamental for \hui{}s: A typical \hui{} consists of multiple \components{} in heterogeneous roles, complementing each other, and forming a single application.
Past and present \hui{}s often consider ``\component{}'' as a synonym for ``device''---creating a cross-device interaction.
% >>> What could it be in future
However, by reconsidering what constitutes an \component{}, we can draw inspiration from concepts such as \textit{complementary interfaces}~\cite{zagermann2022complementary} and related ideas~\cite{elmqvist2023anywhere}:
These concepts are device-agnostic, opening up the research space for combinations of \components{} beyond devices and avoiding being trapped in technological restrictions.
Here, new meaningful combinations of \components{} that include different input and output modalities, combinations of implicit and explicit interaction, and various interaction techniques provide an avenue of research opportunities.
% >>> Final Statement
Thus, we can expect a change in how \hui{}s will be designed, built, and evaluated until they eventually disappear when approaching the ``ultimate display''~\cite{sutherland1965ultimate}.

\section{Conclusion}

We investigated three decades of research in the field of hybrid user interfaces by revisiting their definition, reframing \hui{}s as a fuzzy concept, and discussing future directions.
Despite inconsistent prior usage of the terminology, we identify shared aspects, such as the combination of heterogeneous devices and the merging of visual and interaction spaces.
This combination is especially appealing for mixed reality environments, offering unique opportunities and addressing inherent shortcomings by using a complementary 2D interaction device, and has been extensively explored in prior work with inconsistent terminology.
Our literature survey, therefore, takes a closer look at hybrid user interfaces that combine mixed reality environments with 2D devices and presents a taxonomy of key characteristics.
We see our work as a starting point, enabling the larger community of researchers and practitioners to share their insights and inspire novel systems and interaction techniques.

\section*{Acknowledgements}
We thank all participants of the IEEE ISMAR 2023 Workshop on Hybrid User Interfaces.
%We further thank the reviewers of our initial CHI'25 submission for their valuable feedback.
This work was supported by the \emph{Deutsche Forschungsgemeinschaft} (DFG, German Research Foundation) under Germany’s Excellence Strategy: 
%EXC-2068, 390729961 – Cluster of Excellence ``Physics of Life'' and 
EXC 2050/1, 390696704 – Cluster of Excellence ``Centre for Tactile Internet with Human-in-the-Loop'' (CeTI) of TU Dresden;
by DFG grant 389792660 as part of \href{https://perspicuous-computing.science}{TRR~248 -- CPEC}; 
and by Project-ID~251654672 -- TRR~161.
This work was supported by the German Federal Ministry of Education and Research (BMBF, SCADS22B) and the Saxon State Ministry for Science, Culture and Tourism (SMWK) by funding the competence center for Big Data and AI "ScaDS.AI Dresden/Leipzig";
by the Alexander von Humboldt Foundation (BMBF);
by the Independent Research Fund Denmark (DFF) (grant ID: 10.46540/3104-00008B);
and by Villum Investigator grant VL-54492 by Villum Fonden.

\bibliographystyle{abbrv-doi-hyperref-narrow}
\bibliography{references}

\end{document}